\documentclass[reprint,aps,amsmath,amssymb,superscriptaddress]{revtex4-2}

\usepackage{graphicx}        
\usepackage{dcolumn}        
\usepackage{bm}                
\usepackage{xcolor}           
\usepackage[T1]{fontenc}
\usepackage[hidelinks]{hyperref}


\providecommand{\vect}[1]{{\bf {#1}}}

\newcommand{\Qv}{{\bf Q}}

\newcommand{\kv}{{\bf k}}
\newcommand{\qv}{{\bf q}}

\newcommand{\Qc}{{Q_{\text{c}}}}
\newcommand{\Qcp}{{Q_{\text{c}'}}}

\newcommand{\Hxp}{\hat{H}_{\text{xp}}}

\newcommand{\Ex}{E^{\text{x}}}
\newcommand{\Ep}{E^{\text{p}}}
\newcommand{\Gp}{\Gamma^{\text{p}}}

\newcommand{\vx}{v^{\text{x}}}
\newcommand{\vp}{v^{\text{p}}}
\newcommand{\mx}{m^{\text{x}}}
\newcommand{\mpol}{m^{\text{p}}}

\providecommand{\Gm}[2]{G_{#1,#2}}
\providecommand{\Gr}[2]{G^{\text{ret}}_{#1,#2}}

\providecommand{\average}[1]{\langle {#1} \rangle}
\providecommand{\averageo}[1]{\langle {#1} \rangle_0}

\providecommand{\crea}[1]{\hat{#1}^{\dag}}
\providecommand{\destruye}[1]{\hat{#1}}

\makeatletter

\def\@make@capt@title#1#2{%
  \textbf{#1}\@caption@fignum@sep#2%
}
\makeatother

\begin{document}

\title{Microscopic theory of exciton polarons in transition metal dichalcogenides}

\author{Carlos Maciel-Escudero}
\email{macieles@uni-marburg.de}
\affiliation{Department of Physics, Philipps-Universit\"{a}t Marburg, 35032 Marburg, Germany} 
\affiliation{mar.quest|Marburg Center for Quantum Materials and Sustainable Technologies, 35032 Marburg, Germany}
\author{Giuseppe Meneghini} 
\affiliation{Department of Physics, Philipps-Universit\"{a}t Marburg, 35032 Marburg, Germany} 
\affiliation{mar.quest|Marburg Center for Quantum Materials and Sustainable Technologies, 35032 Marburg, Germany}
\author{Jamie M. Fitzgerald} 
\affiliation{Department of Physics, Philipps-Universit\"{a}t Marburg, 35032 Marburg, Germany} 
\affiliation{mar.quest|Marburg Center for Quantum Materials and Sustainable Technologies, 35032 Marburg, Germany}
\author{Ermin Malic} 
\affiliation{Department of Physics, Philipps-Universit\"{a}t Marburg, 35032 Marburg, Germany} 
\affiliation{mar.quest|Marburg Center for Quantum Materials and Sustainable Technologies, 35032 Marburg, Germany}

\begin{abstract}
Polarons are composite quasiparticles formed when charge carriers strongly couple to lattice vibrations, manifesting as carriers dressed by a surrounding cloud of phonons. Although this dressing fundamentally renormalizes the effective mass and mobility of carriers, its impact on electrically neutral exciton polarons in 2D semiconductors has received limited attention. In particular, despite the growing interest in polaronic effects in atomically thin semiconductors, only a few theoretical studies have addressed spectral signatures and renormalized properties of exciton polarons in these materials. Here, we develop a microscopic model to describe exciton polarons in transition metal dichalcogenides. Applying this model to $1s$ acoustic polarons in an MoSe$_2$ monolayer, we reveal a distinct kink in the polaron dispersion, accompanied by a pronounced broadening of its spectral function. These spectral signatures mark a critical transition from a regime of well-defined bound exciton polarons, characterized by a renormalized energy, reduced group velocity, and enhanced effective mass, to a regime dominated by exciton-phonon scattering. Furthermore, by incorporating intervalley exciton-phonon interaction, we demonstrate that the energetic position of momentum-dark excitons strongly modifies the exciton-polaron lifetime. Overall, our work  provides a microscopic framework for determining renormalized properties of exciton polarons and establishes experimentally accessible polaron signatures in 2D semiconductors.
\end{abstract}

\maketitle

\section*{\label{intro} Introduction}

Semiconducting transition metal dichalcogenide (TMD) monolayers host excitons, which are tightly bound electron-hole pairs \cite{Malic2018,Bernhard2018}.  The low dimensionality of TMD monolayers results in a reduced dielectric screening and strong  geometrical confinement, which substantially enhance the Coulomb interaction between electrons and holes. Consequently, excitons in TMD monolayers exhibit exceptionally large binding energies and oscillator strengths \cite{Chernikov2014,Shan2014}, enabling the formation of excitons that remain stable even at room temperature. Owing to these properties, excitons dominate the optical and optoelectronic response of TMD monolayers, governing their absorption and photoluminescence \cite{Heinz2010,Strano2012,Heinz2014,Li2014,Yu2015,Heinz2015,Ermolaev2020,Malic2022}. Beyond their central role in optical spectroscopy, excitons in TMD monolayers have shown the potential to form a variety of many-body quasiparticle complexes, including exciton-polaritons \cite{Menon2015,Schneider2016,Schneider2018,Fitzgerald2022}, charged excitons (trions or Fermi polarons) \cite{Mak2013,Ross2013,Plechinger2016,Courtade2017,Arora2019,Malic2024}, and biexcitons \cite{You2015,Steinhoff2018,Barbone2018}. In this context, and among the many-body interactions determining exciton dynamics, coupling to phonons is particularly important, as it influences the relaxation process \cite{Knorr2017,Brem2018,Schmitt2022}, coherence lifetime \cite{Moody2015,Selig2016,Dey2016}, and transport \cite{Malic2023,Malic2019,Meneghini2025} of excitons. When this coupling is sufficiently strong, excitons can be dressed by lattice distortions, forming exciton polarons.

%
\begin{figure}[!ht]
\begin{center}
\includegraphics[width=0.9\linewidth]{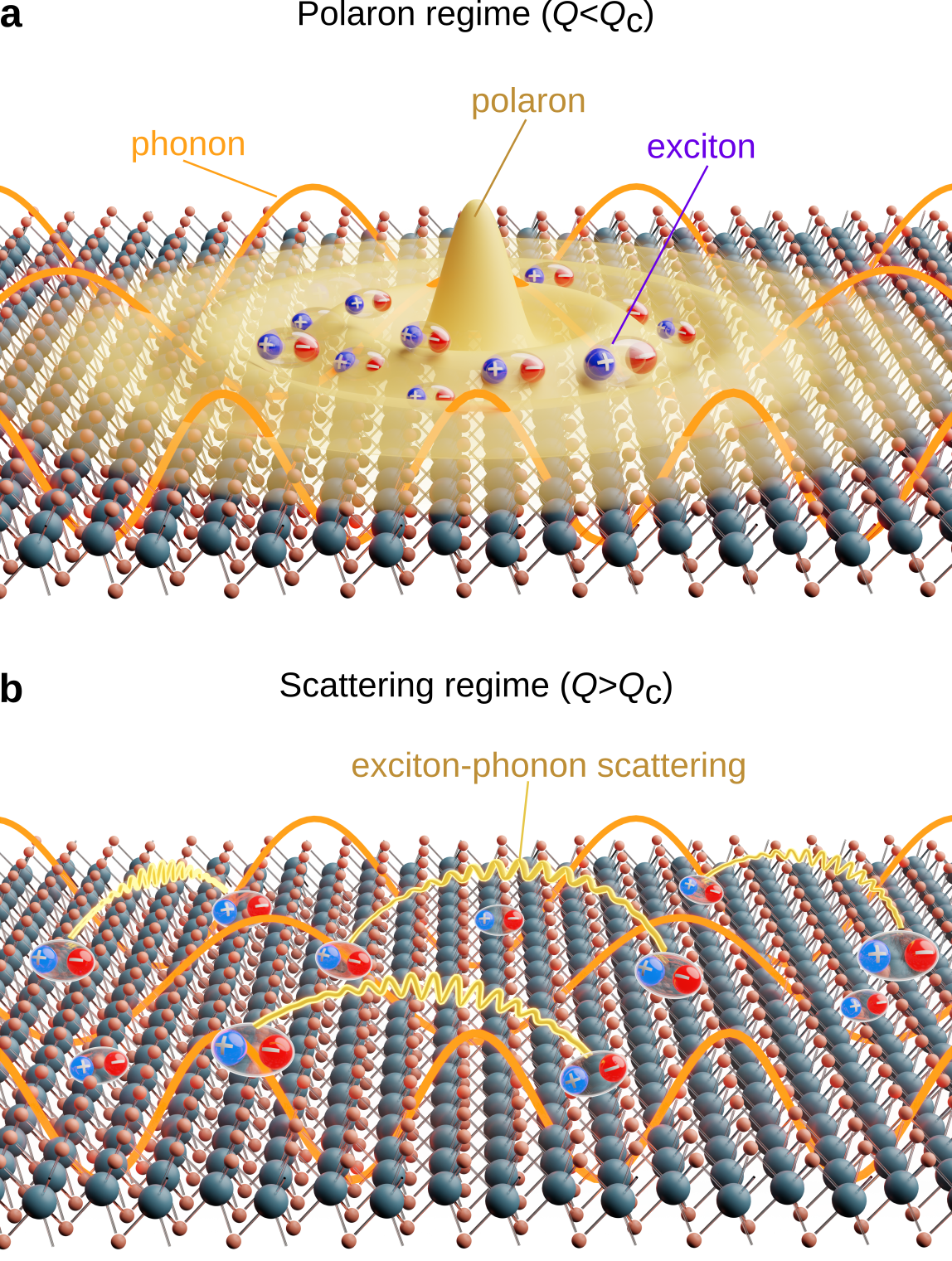}
\caption{\textbf{Schematic illustration of the exciton-polaron and exciton-phonon scattering regimes.}  \textbf{a} Excitons (bound electron-hole pairs, blue and red spheres) interact with the TMD crystal lattice. This strong interaction with lattice vibrations (phonons, orange waves) effectively dresses excitons with a phonon cloud, forming exciton polarons (yellow wavepacket) below a critical momentum $\Qc$. \textbf{b} Above $\Qc$, as excitons propagate through the crystal lattice, they can experience discrete scattering events with phonons (yellow curves). Here, excitons absorb or emit phonons, altering their center-of-mass momentum.}
\label{fig1}
\end{center}
\end{figure}
%

Polarons have been extensively studied in the context of charge carriers in semiconductors  \cite{AlexandrovBook2007,EminBook213,Giustino2017,Franchini2021}. These quasiparticles arise when an electron traveling through the semiconductor induces a local distortion within the crystal lattice, thus displacing the constituent ions from their respective equilibrium positions. This distortion produces a local polarization in the lattice that acts as a potential well, effectively binding the electron to the induced lattice deformation. This process can be conceptually described as the electron being dressed by the surrounding cloud of phonons that accompany its motion through the crystal. As a consequence of this dressing, the properties of the electron are modified, resulting in an energy redshift, a reduced group velocity, and an enhanced effective mass of polarons compared to that of free electrons with the same momentum \cite{Franchini2021}. 

Polaron effects have been widely observed across a variety of material systems, including organic semiconductors \cite{Coropceanu2007}, halide perovskites \cite{Tretiak2020}, and 2D materials \cite{Kang2018,Sio2023,Knorr2024}. Their emergence can lead to modifications of charge carrier mobility \cite{Coropceanu2007}, effective mass \cite{Chen2015}, surface reactivity \cite{Reticcioli2019}, among other properties. Although traditional polarons formed by electron-phonon interactions have been extensively studied in conventional semiconductors \cite{AlexandrovBook2007,EminBook213,Franchini2021}, achieving a comprehensive understanding of electrically neutral exciton polarons in 2D semiconductors remains an active area of research. Recent theoretical advancements have enabled first-principles density functional theory (DFT) calculations of exciton polarons \cite{Giustino2024}. However, these approaches are computationally demanding for excitonic systems. Furthermore, despite the extensive evidence showing that intervalley scattering is crucial for understanding the optical response and exciton relaxation in TMDs \cite{Selig2016,Knorr2017,Gunnar2018,Lindlau2018,Bae2022}, its impact on exciton-polaron physics remains unexplored. Consequently, an accessible yet robust theoretical framework capable of microscopically capturing the renormalized quasiparticle properties and spectral signatures of exciton polarons in TMDs is still pending.

In this work, we develop a microscopic theoretical framework based on the Matsubara Green’s function formalism \cite{Walecka2003,Bruus2004,Nolting2009} that is capable of describing exciton-polaron quasiparticles in TMD monolayers. Applying this framework to investigate exciton polarons in an exemplary MoSe$_2$ monolayer, and considering a $1s$ exciton state interacting with acoustic phonons, we find that the dispersion relation obtained from the Dyson equation exhibits a distinct kink at a characteristic, critical momentum $\Qc$. The appearance of this kink is related to a broadening of the spectral function, similar to previous observations in electron-polaron systems \cite{Schrieffer1963,Echenique2009}. This feature serves as a signature of the presence of exciton polarons in TMD monolayers and highlights the crossover between a regime of bound exciton polarons and a regime dominated by exciton-phonon scattering, as schematically shown in Fig. \ref{fig1}. Below $\Qc$, well-defined bound polarons are formed, characterized by an energy redshift, a reduced group velocity, and a mass enhancement compared to that of $1s$ excitons with the same center-of-mass momentum. We further demonstrate that as the temperature increases from cryogenic values to room temperature, the enhanced thermal phonon population induces an energy redshift by more than 10\,meV and increases the exciton-polaron mass by up to 30\%. Finally, by extending our methodology to incorporate intervalley momentum-dark $\text{KK}'$ excitons, we demonstrate that intervalley phonon scattering reduces the lifetime of $\text{KK}$ exciton polarons in monolayer WSe$_2$, whereas the lifetime in MoSe$_2$ remains largely unaffected. Thus, our findings establish intervalley scattering as a critical mechanism determining exciton-polaron dynamics in TMDs.

\section*{\label{results}  Results}

\subsection*{Microscopic model}

We describe the interacting system of excitons and phonons within a TMD monolayer using the low-density excitonic Hamiltonian \cite{Malic2020}:
\begin{align}
\label{Hamiltonian}
\Hxp &= \sum_{\mu, \Qv} \Ex_{\mu \Qv}\crea{X}_{\mu \Qv} \destruye{X}_{\mu \Qv} + \sum_{\alpha, \kv} \hbar \Omega_{\alpha \kv}\,  \crea{b}_{\alpha \kv} \destruye{b}_{\alpha \kv}  \\
\nonumber
&+\sum_{\mu \nu \Qv, \alpha \qv} \mathcal{D}_{\alpha \qv}^{{\nu \mu}} \crea{X}_{\mu \Qv + \qv} \destruye{X}_{\nu \Qv} (\destruye{b}_{\alpha \qv} + \crea{b}_{\alpha, -\qv}),
\end{align}
where $\crea{X}_{\mu \Qv}$ and $\destruye{X}_{\mu\Qv}$ are the bosonic creation and annihilation operators for an exciton in the state $\mu$ with the center-of-mass momentum $\Qv$.  The operators $\crea{b}_{\alpha \qv}$ and $\destruye{b}_{\alpha \qv}$ are the corresponding bosonic creation and annihilation operators for a phonon in the mode $\alpha$ with momentum $\qv$. The terms $\Ex_{\mu \vect{Q}}$ and $\hbar \Omega_{\alpha \qv}$ represent the exciton and phonon dispersion relations, respectively. The first is obtained from the Wannier equation combined with the effective mass approximation, while the latter is described using Einstein and Debye models.  The last term in Eq. \eqref{Hamiltonian} describes the scattering of an exciton from the state $(\nu,\Qv)$ to the state $(\mu, \Qv + \qv)$ by emitting or absorbing a phonon, weighted by the exciton-phonon matrix elements $\mathcal{D}_{\alpha \qv}^{{\nu \mu}}$. These matrix elements are modeled within the deformation potential approximation \cite{Kim2014,Brem2018}.
 
To describe the formation of exciton polarons within a TMD monolayer, we use both the Matsubara Green's function and equation-of-motion formalisms \cite{Walecka2003,Bruus2004,Nolting2009}. Taking the many-particle Hamiltonian from Eq. \eqref{Hamiltonian}, we obtain the renormalized polaron energies by calculating the self-energy $\Sigma_{\mu\nu\Qv}$. This quantity contains crucial information regarding the complex polaron energy $\Ep_\Qv$, whose real component determines the dispersion relation of exciton polarons, and whose imaginary component is related to the finite lifetime of the quasiparticle. In this work, we restrict our analysis to the Fan-Migdal self-energy \cite{Giustino2017}, which represents the first non-zero term in the perturbative expansion of the exciton-phonon propagator. Physically, this approximation describes single-phonon processes where an exciton emits and subsequently reabsorbs a phonon.  Here, the renormalized polaron energies are determined by the poles of the retarded Green's function $\Gr{\mu\Qv}{\nu\Qv}$, which are obtained by solving the following secular equation derived from the Dyson equation: 
\begin{equation}
\label{Dyson}
\text{det}\left[\left(\Ep_\Qv  - \Ex_{\mu \Qv} \right)\delta_{\mu\nu} - \Sigma_{\mu\nu\Qv}(\Ep_\Qv)  \right] = 0.
\end{equation}
 Further details regarding the derivation of Eq. \eqref{Dyson} and the numerical calculations are provided in Methods and Supplementary Material.

\subsection*{Signatures of exciton polarons in TMD monolayers}

We apply the developed model to MoSe$_2$ monolayers as a representative TMD material. Molybdenum-based TMDs host excitons with binding energies on the order of hundreds of meV \cite{Heinz2015,Rivera2015} and are characterized by large exciton-phonon matrix elements \cite{Knorr2017}, making them ideal material platforms for exploring exciton-polaron physics. Specifically, we focus on polarons formed by $1s$ excitons composed of electrons and holes localized near the high-symmetry K points of the electronic band structure, interacting with both transverse (TA) and longitudinal (LA) acoustic phonon modes of the TMD monolayer. We restrict our analysis to acoustic phonons since they provide the main contribution to exciton–phonon scattering at cryogenic temperatures. The methodology can, however, equally describe interactions with optical phonons (Supplementary Note 2), which become increasingly relevant at elevated temperatures.

%
\begin{figure}[t!]
\begin{center}
\includegraphics[width=0.9\linewidth]{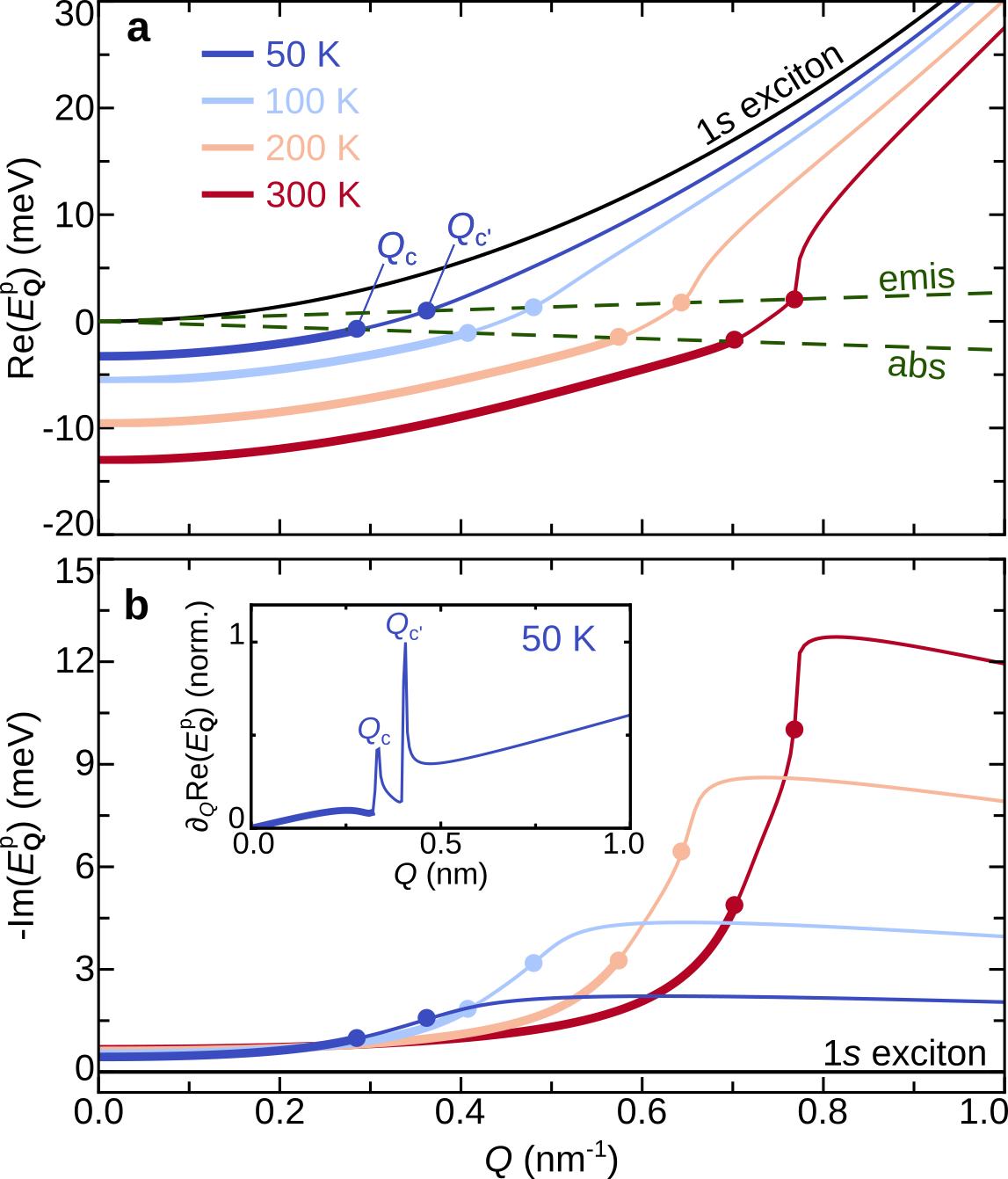}
\caption{\textbf{Formation of bound exciton polarons.} \textbf{a} Real component of the exciton polaron energy, $\text{Re}(\Ep_\Qv)$, for an MoSe$_2$ monolayer as a function of the magnitude of the center-of-mass momentum $Q$ at different temperatures. The black curve represents the bare $1s$ exciton dispersion, $\Ex_{1s,\Qv}$. The green dashed lines show $\Ex_{1s,\Qv=\bf 0} \pm \hbar\Omega_{\text{TA/LA},\Qv}$, where $\hbar\Omega_{\text{TA/LA},\Qv} = \hbar v_{\text{TA/LA}}Q$ is the dispersion of the TA/LA acoustic phonons. The upper ($+$) and lower ($-$) branches correspond to phonon emission (emis) and absorption (abs), respectively. Their intersections with the polaron dispersions are indicated by the color dots and mark the energy conservation condition $\text{Re}(\Ep_\Qv)=\Ex_{1s,\Qv=\bf 0} \pm \hbar\Omega_{\text{TA/LA},\Qv}$ at the critical momenta $Q_{\text{c}}$ and $Q_{\text{c}'}$ for each temperature. \textbf{b} Negative imaginary component of the polaron energy, $-\text{Im}(\Ep_\Qv)$, as a function of $Q$ across the same temperatures as in panel a. The inset shows the derivative of $\text{Re}(\Ep_\Qv)$ with respect to $Q$ for the polaron dispersion at $T=50\,\text{K}$, normalized to its maximum value. Note that the energy axes are defined relative to the exciton energy at $\Qv={\bf 0}$, i.e., $\Ep_\Qv -  \Ex_{1s, \vect{Q=0}}$.}
\label{fig2}
\end{center}
\end{figure}
%

Figure \ref{fig2}a shows the real component of the exciton polaron dispersion, $\text{Re}(\Ep_\Qv)$, obtained by solving the Dyson equation (Eq. \eqref{Dyson}) for an MoSe$_2$ monolayer at different temperatures. For comparison, we also plot the dispersion of the bare $1s$ exciton ($\Ex_{1s, \vect{Q}}$, black line). We observe that the polaron dispersion is redshifted by about 10 meV depending on the temperature. More importantly, two distinct regimes emerge, indicated by the varying thickness of the lines. The onset of the transition between these regimes is marked by the appearance of a small kink in the polaron dispersion, indicated by a colored dot. This feature occurs precisely at the momentum $\Qc$ where the exciton polaron energy matches the energy of either a TA or a LA phonon, satisfying the energy conservation condition $\text{Re}(\Ep_{\Qv_\text{c}}) = \Ex_{1s, \vect{Q=0}} - \hbar \Omega_{\text{TA/LA},{\Qv_\text{c}}}$. This implies that the microscopic origin of the kink is related to the opening of phonon absorption scattering channel.

To verify this, we plot the energy $\Ex_{1s, \vect{Q=0}} - \hbar \Omega_{\text{TA/LA},\Qv}$, where $\hbar \Omega_{\text{TA/LA},\Qv}= \hbar v_\text{TA/LA} Q$ is the linear dispersion of the TA and LA phonons, and  $v_\text{TA/LA}$ is the effective sound velocity of each acoustic mode \cite{Kim2014}. In monolayer MoSe$_2$, both the TA and LA phonons propagate with the same effective velocity (see Supplementary Note 2), thus $\Ep_{1s,\Qv=0} - \hbar\Omega_{\text{TA},\Qv}$ and $\Ep_{1s,\Qv=0} - \hbar\Omega_{\text{LA},\Qv}$ coincide and are shown as a single curve in Fig. \ref{fig2}a (green dashed line labeled ``abs''). Notably, the position of the kink at $\Qc$ matches exactly with the intersection between the exciton-polaron dispersion and the green dashed line, confirming the connection between the appearance of a kink and the opening of the phonon absorption channel. As the temperature increases, the exciton polaron dispersion shifts to lower energies, and the position of the kink $\Qc$ shifts to higher momenta. To resolve the small kink more clearly within the exciton-polaron dispersion, we reduce the numerical broadening $\eta$ in Eq. \eqref{Dyson} from 2\,meV to 0.2\,meV and compute the derivative of $\text{Re}(\Ep_\Qv)$ with respect to $Q$ at $T = 50\,\text{K}$ (inset of Fig. \ref{fig2}b). The kink then appears as a sharp peak at $\Qc$.

Importantly, the kink separates the regime of bound exciton-polarons from the exciton-phonon scattering regime above the critical momentum $\Qc$. The transition between these two regimes becomes also evident in the negative imaginary component of the complex polaron energy, $\Gp_\Qv=-\text{Im}(\Ep_\Qv)$, shown in Fig. \ref{fig2}b. Physically, this quantity determines the exciton-phonon scattering rate and the linewidth of the polaron. It is inversely proportional to the polaron lifetime as $\tau_\Qv=\hbar/(2\Gp_\Qv)$ \cite{Giustino2017,Louie2023}. For momenta significantly below $\Qc$, the exciton polaron exhibits a negligible $\Gp_\Qv$ (indicated by thick colored lines close to zero in Fig. \ref{fig2}b), corresponding to an infinite polaron lifetime. The small non-zero value of $\Gp_\Qv$ close to $Q=0$ is a consequence of the finite numerical broadening introduced by $\eta$ in Eq. \eqref{Dyson}, which also smoothens the onset of $\Gp_\Qv$ at $\Qc$ (Supplementary Note 2). For $Q<\Qc$, the exciton polaron manifests as a well-defined bound state, corresponding to a real zero in Eq. \eqref{Dyson} and a real pole in the retarded Green's function (Eq. \eqref{Green2}). Conversely, for momenta above $\Qc$, the polaron acquires sufficient energy to absorb an acoustic phonon, resulting in a finite $\Gp_\Qv$ and, consequently, a finite polaron lifetime. Here, the description in terms of a well-defined quasiparticle breaks down as $\Ep_\Qv$ evolves into a complex pole, and the system transitions into the exciton-phonon scattering continuum. Even in this regime, excitons strongly interact with  phonons, exhibiting distinct polaronic effects, such as an energy redshift. 

In analogy, the second kink at $\Qcp > \Qc$, visible in both $\text{Re}(\Ep_\Qv)$ (Fig. \ref{fig2}a) and $\Gp_\Qv$ (Fig. \ref{fig2}b), emerges when the exciton polaron has enough energy to emit a TA or LA phonon, i.e., when the phonon emission channel opens. To verify this, we plot $\Ex_{1s,\Qv=0} + \hbar\Omega_{\text{TA/LA},\Qv}$ in Fig. \ref{fig2}a (green dashed line labeled ``emis''). As for absorption, the TA and LA contributions coincide and are shown as a single dashed line. The kink appears exactly at the momentum $\Qcp$ where the green line intersects the exciton-polaron dispersion, confirming that it originates from phonon emission. Similar to the first kink, it also produces a sharp peak in the derivative of $\text{Re}(\Ep_\Qv)$ (inset of Fig.~\ref{fig2}b). Both kinks, at $\Qc$ and $\Qcp$, thus mark the momenta at which phonon scattering channels open. Mathematically, the kinks arise where the polaron energy $\Ep_\Qv$ reaches a branch point of the retarded Green’s function (Eq.~\eqref{Green2}). This connection is discussed in detail in Supplementary Note 3.

%
\begin{figure}[!ht]
\begin{center}
\includegraphics[width=0.9\linewidth]{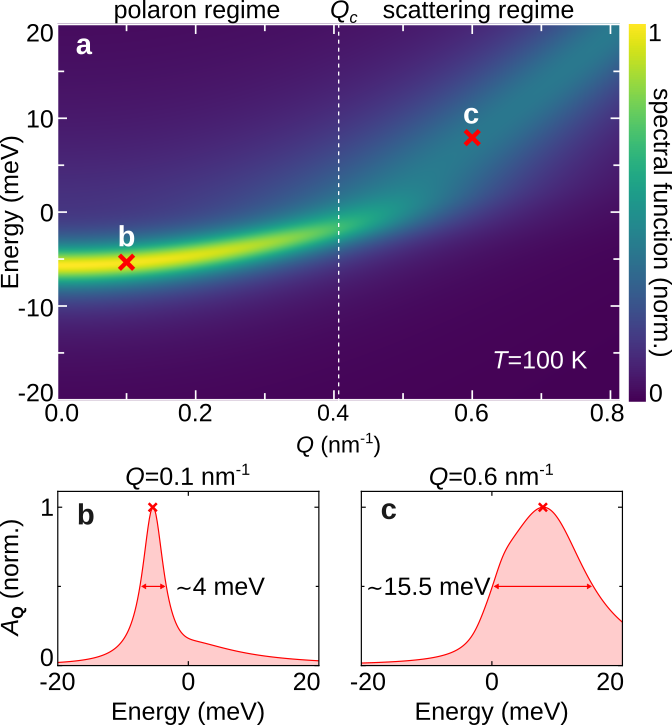}
\caption{\textbf{Momentum-resolved spectral function of an MoSe$_2$ monolayer.} \textbf{a} Normalized 2D contour plot of the spectral function, $A_{\Qv}(E)$, as a function of center-of-mass momentum $Q$ and real energy $E$ evaluated at 100\,K. The spectral function is calculated using a fixed numerical broadening of $\eta=2\,\text{meV}$ while sweeping the real part of the energy, $E$, along the vertical axis. The vertical dashed line indicates the position of the critical momentum $\Qc=0.407\,\text{nm}^{-1}$. \textbf{b}-\textbf{c} Normalized spectral functions obtained at $Q=0.1\,\text{nm}^{-1}$ (panel b) and $Q=0.6\,\text{nm}^{-1}$ (panel c), corresponding to the positions marked by the red crosses in panel a. The double-headed arrows indicate the FWHM. The energy axes in all panels are defined relative to $\Ex_{1s, \vect{Q=0}}$. }
\label{fig3}
\end{center}
\end{figure}
%

To examine the physical consequences of the transition between the bound exciton-polaron and the exciton-phonon scattering regime, we calculate the spectral function $A_\Qv(E)= -\text{Im}[1/(E - \Ex_{1s,\Qv} - \Sigma_{1s,1s,\Qv}(E) +i\eta)]/\pi$. This quantity represents the momentum-resolved density of states (DOS) of the system \cite{Snoke2000}. For the calculation of the spectral function, we use a fixed numerical broadening parameter $\eta=2\,\text{meV}$. As shown in Fig. \ref{fig3}a at $T=100\,\text{K}$, the spectral function exhibits a well-defined peak (yellow region) that monotonically shifts to higher energies as $Q$ increases up to $\Qc \approx 0.4\,\text{nm}^{-1}$ (vertical dashed line). The appearance of this peak is attributed to the formation of bound exciton polarons within the monolayer, consistent with the emergence of a well-defined pole of the retarded Green's function (Eq. \eqref{Green2}) on the real energy axis, i.e., a real solution of Eq. \eqref{Dyson}  at momenta smaller than $\Qc$ (Fig. \ref{fig2}). Above $0.4\,\text{nm}^{-1}$, the peak in Fig. \ref{fig3}a continues shifting to higher energies but broadens significantly due to the dominant scattering between excitons and acoustic phonons. In Figs. \ref{fig3}b and \ref{fig3}c, we show cuts of the spectral function at fixed momenta of $Q=0.1\,\text{nm}^{-1}$  and $Q=0.6\,\text{nm}^{-1}$. The deviation of the spectral function $A_\Qv(E)$ from a symmetric Lorentzian line shape in panels b and c is a consequence of the energy-dependent shift introduced by the self-energy $\Sigma_{1s,1s,\Qv}(E)$ \cite{Knorr2017}. More importantly, the broadening of the spectral function is clearly reflected in the full-width at half-maximum (FWHM) of both spectra. This quantity is related to the imaginary part of the polaron energy through $\text{FWHM} = 2 (\eta + \Gp_\Qv)$ (see Supplementary Note 2), which allows us to extract $\Gp_\Qv$ and thus estimate the polaron lifetime $\tau_\Qv$ directly from the spectral function. Below $Q_{\text{c}}$ (panel b in Fig. \ref{fig3}), the spectrum exhibits a FWHM of about 4\,meV, which equals $2\eta$, the width set by the numerical broadening, i.e., $\Gp_\Qv$ vanishes in this regime and thus bound exciton polarons are long-lived, with a lifetime not limited by phonon scattering, in agreement with the solution of the Dyson equation shown in Fig. \ref{fig2}b. In stark contrast, the spectrum above $Q_{\text{c}}$ (Fig. \ref{fig3}c) broadens to a FWHM of $15.5\text{ meV}$ due to the opening of exciton–phonon scattering channels, which reduce the polaron lifetime to about 57\,fs, for the temperature and momentum shown in Fig. \ref{fig3}c. The FWHM extracted from the spectral function is in good agreement with the values obtained from experimental optical absorption spectra of $\text{MoSe}_2$ at $T = 100\,\text{K}$ \cite{Knorr2017}.

Note that the broadening and kink features appearing in the spectral function have already been observed in the context of electron polarons in angle-resolved photoemission spectroscopy (ARPES) measurements of oxide materials \cite{Moser2013,Verdi2017}. In these experiments, the spectral function characterizes the momentum-resolved single-particle excitation spectrum. We thus envision that the emergence of analogous features in the excitonic spectral function may provide experimentally accessible signatures of exciton polarons in TMD monolayers. In addition, exciton-polaron formation is also expected to substantially modify the properties of underlying excitonic quasiparticles. We therefore now examine the renormalization of exciton energy, group velocity, and effective mass.

%
\begin{figure}[!t]
\begin{center}
\includegraphics[width=0.9\linewidth]{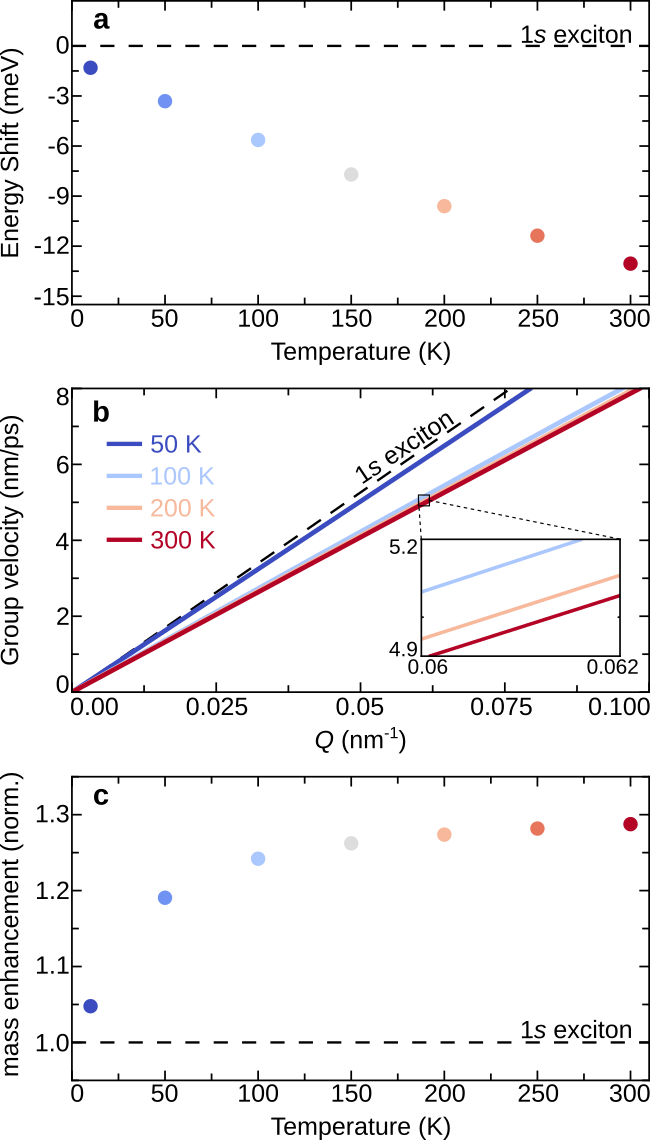}
\caption{\textbf{Polaron-induced change in exciton properties.} \textbf{a} Energy shift $\Delta E = \Ex_{1s,\Qv=0} - \text{Re}(\Ep_{\Qv=0})$ of the exciton polaron relative to the bare $1s$ exciton at $Q = 0$, as a function of temperature. \textbf{b} The corresponding polaron group velocities, $\vp_\Qv$, in direct comparison to the 1s exciton  group velocity, $\vx_{1s,\Qv}$ denoted by the black dashed line. The inset shows a zoom of the region $0.06\,\text{nm}^{-1} \leq Q \leq 0.062\,\text{nm}^{-1}$ and $4.9\,\text{nm/ps} \leq \vp_\Qv \leq 5.2\,\text{nm/ps}$, highlighting the reduction of $\vp_\Qv$ with increasing temperature. \textbf{c} Temperature dependence of the polaron mass enhancement, $\mpol/\mx_{1s}$, obtained around $Q=0$. }
\label{fig4}
\end{center} 
\end{figure}
%

\subsection*{Polaron-induced renormalization of exciton properties}

Phonon dressing renormalizes the properties of the bare exciton. To quantify this, we focus on the low-momentum region around $0<Q<0.1\,\text{nm}^{-1}$ and extract the temperature-dependent polaron energy shift, $\Delta E$, group velocity, $\vp_\Qv$, and effective mass, $\mpol$, from the exciton-polaron dispersions shown in Fig. \ref{fig2}a. All quantities are compared with those of the bare $1s$ exciton.
In Fig.\ref{fig4}a, we plot the polaron redshift at $Q = 0$ relative to the bare exciton energy (horizontal dashed line). We find that at cryogenic temperatures ($T = 10\,\text{K}$, blue dot), the polaron energy is redshifted by about 1.5\,meV. This shift increases substantially with temperature, reaching $\sim13\,\text{meV}$ at 300\,K. This is in a good agreement with the experimentally extracted temperature dependence of the polaron energy shift in MoSe$_2$ monolayers \cite{Knorr2017}. The predicted values are smaller, which can be  attributed to the optical phonons that have not been included in our simulations. 

We further find that the group velocity $\vp_\Qv$ is strongly renormalized (see Fig. \ref{fig4}b), resulting in a reduced velocity relative to that of the bare $1s$ exciton. Note that in the low-momentum region, the polaron group velocity $\vp_\Qv$ increases linearly with momentum at all temperatures, reflecting a parabolic polaron dispersion around its minimum at $Q = 0$, that is, $\text{Re}(\Ep_\Qv) \approx \text{Re}(\Ep_{\Qv=0}) + \hbar^2Q^2/(2\mpol)$. The larger the temperature, the smaller the slope becomes and the change quickly saturates for $T >\,100\,\text{K}$ (inset of Fig. \ref{fig4}b). This has a direct impact on the effective polaron mass $\mpol$. Figure \ref{fig4}c compares $\mpol$  with the bare exciton mass, $\mx_{1s}$ and we find that the polaron mass is enhanced by $\sim5\%$ at $10\,\text{K}$ and this enhancement increases monotonically with temperature, reaching a saturation value of $\sim30\%$ at room temperature. These theoretical predictions can be validated via magneto-optical spectroscopy experiments \cite{Paulina2024}, with the reduced group velocity and mass renormalization having direct implications on exciton-polaron transport properties.

%
\begin{figure*}[!t]
\begin{center}
\includegraphics[width=0.9\linewidth]{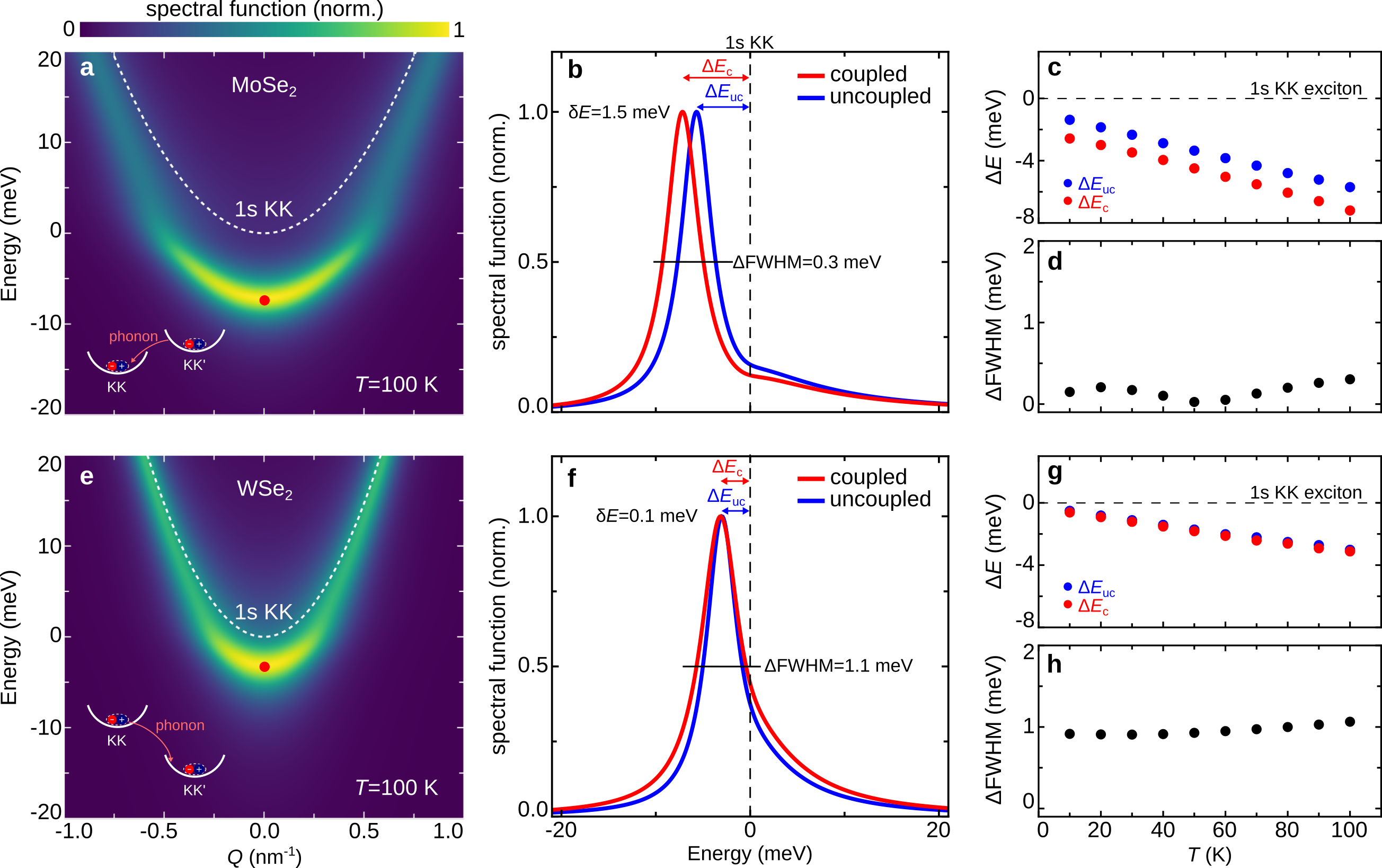}
\caption{\textbf{Spectral functions of MoSe$_2$ and WSe$_2$ including intervalley scattering.} \textbf{a} Spectral function $A_\Qv(E)$ of an MoSe$_2$ monolayer at $T=100\,\text{K}$ as a function of the center-of-mass momentum $Q$ and energy $E$. The calculation considers intervalley coupling between the $1s\text{ KK}$ and $1s \text{ KK}'$ excitons via acoustic K-point phonon scattering. The white dashed curve indicates the dispersion of the $1s \text{ KK}$ exciton, $\Ex_{1s\text{KK}, \Qv}$. \textbf{b} Normalized spectral functions at $Q=0$ with (red) and without (blue) intervalley coupling. The vertical dashed line marks $\Ex_{1s\text{KK},\Qv=\bf 0}$ and the doubled-headed arrows indicate the energy shift $\Delta\text{E}$ of each peak maximum relative to $\Ex_{1s\text{KK},\Qv=\bf 0}$. The energy difference between the peak maxima of the coupled and uncoupled spectra is denoted by $\delta E = |\Delta E_\text{c} - \Delta E_\text{uc}|$. The horizontal black line indicates the linewidth difference $\Delta\text{FWHM}$ between the two spectra. \textbf{c} Polaron energy redshift $\Delta E$ as a function temperature for the coupled (red dots) and uncoupled (blue dots) systems, referenced to $\Ex_{1s\text{KK}, \Qv=\bf 0}$ indicated by the horizontal dashed line. \textbf{d} Linewidth difference $\Delta\text{FWHM}$ as a function of temperature. \textbf{e}-\textbf{h} Same as \textbf{a}-\textbf{d} but for a WSe$_2$ monolayer. The red dots in the 2D contour plots mark the peak at $Q = 0$, while the insets illustrate the coupling to intervalley excitons via scattering of phonons. All spectra are normalized to their peak maximum and the energy axes are defined relative to $\Ex_{1s\text{KK}, \Qv=\bf 0}$ of the respective material.}
\label{fig5}
\end{center}
\end{figure*}
%

\subsection*{Intervalley coupling and exciton-polaron lifetime}

Thus far, we have focused on polarons in MoSe$_2$ formed via the coupling between $1s$ intravalley KK excitons and intravalley acoustic phonons. We now expand our theoretical framework to incorporate intervalley excitons. Specifically, we consider intervalley $\text{KK}'$ excitons composed of electrons and holes localized at the $\text{K}'$ and K points of the Brillouin zone, respectively. The scattering of K-point phonons mediate the exciton-phonon interaction of KK excitons into these $\text{KK}'$ states, significantly altering both the coherence lifetime and the linewidth of excitonic resonances \cite{Selig2016,Bae2022}. To capture the influence of intervalley coupling on the formation of exciton polarons, we incorporate the intervalley degrees of freedom into the retarded Green's function, $\Gr{\mu\Qv}{\nu\Qv}$, and compute the spectral function as $A_{\Qv} = -\text{Im}[\text{Tr}(\Gr{\mu\Qv}{\nu\Qv})]/\pi$, where the indices $\mu$ and $\nu$ run over the $1s\text{ KK}$ and $1s\text{ KK}'$ excitonic states \cite{Huang2023}. As discussed in the context of Fig. \ref{fig3}, evaluating the spectral function yields information comparable to solving the full Dyson equation, as it directly provides the polaron spectrum, including the renormalized resonance energies and an estimate of their lifetimes.  In the following, we thus use the spectral function to determine modifications in the exciton-polaron properties due to intervalley coupling. 

To elucidate the impact of intervalley coupling, we compare the spectral function of $\text{MoSe}_2$, where the ground state corresponds to momentum-bright $\text{KK}$ excitons, with that of $\text{WSe}_2$, where the ground state is given by momentum-dark $\text{KK}'$ excitons. Figure \ref{fig5}a shows the normalized 2D contour map of the spectral function for the $1s\text{ KK}$ exciton in MoSe$_2$ in the presence of coupling to the $1s\text{ KK}'$ exciton. For comparison, we also plot the dispersion relation of the bare $1s\text{ KK}$ exciton (white dashed line). From Fig. \ref{fig5}a, we observe that the contour map features a prominent peak in the low-momentum region, corresponding to the $1s\text{ KK}$ exciton polaron. At higher momenta, this peak broadens due to intravalley exciton-phonon scattering, analogous to the behavior shown in Fig. \ref{fig3}a. For completeness, the 2D contour maps of the spectral functions of both the $1s\text{ KK}$ and $1s\text{ KK}'$ exciton polarons are provided in Supplementary Note 4. In monolayer MoSe$_2$, the energy of the optically dark $1s\text{ KK}'$ exciton state lies approximately $10\text{ meV}$ above that of the bright $1s\text{ KK}$ exciton \cite{Selig2018}, as illustrated in the inset of Fig. \ref{fig5}a. This energy offset is insufficient to activate intervalley scattering via emission of K-point acoustic phonons, as TA and LA phonons at the K point possess higher energies of approximately $16.6\,\text{meV}$ and $19.9\,\text{meV}$, respectively \cite{Kim2014}. Scattering into the $\text{KK}'$ valley therefore requires the absorption of a K-point phonon, which is suppressed by their low thermal occupation. Intervalley coupling thus redshifts the $1s\text{ KK}$ polaron by only $\sim 1.5\,\text{meV}$. This is illustrated in Fig. \ref{fig5}b, which shows slices of the spectral function at $Q = 0$ around the $\text{KK}$ polaron resonance, without (blue) and with (red) coupling to the $\text{KK}'$ exciton.

To verify that the $1s\text{ KK}$ exciton-polaron lifetime is unaltered by the intervalley coupling, we analyze the FWHM of the spectral functions at $Q=0$ with and without coupling. In both cases, we find  a FWHM of approximately 4\,meV, which is determined by the numerical broadening parameter $\eta$ = 2\,meV used in the calculation of the spectral function. The minor linewidth difference of $\Delta\text{FWHM} = 0.3\,\text{meV}$ between the two spectra is attributed to lineshape modifications from adjacent spectral features, such as phonon sidebands introduced by the self-energy \cite{Knorr2017}, rather than active intervalley scattering. To confirm that phonon-mediated intervalley scattering remains thermally inactive when varying temperature and that intervalley coupling induces purely an energy redshift on the polaron, we calculate the spectral functions for both the uncoupled and coupled systems for temperatures between $10\,\text{K}$ and $100\,\text{K}$ (Supplementary Note 5). The extracted peak energies and $\Delta\text{FWHM}$ are shown in Figs. \ref{fig5}c and \ref{fig5}d, respectively. Across this entire temperature range, the polaron energy shift (red points in panel c) decreases monotonically, reaching $\sim8\,\text{meV}$ at $T=100\,\text{K}$, consistently lying below the uncoupled reference (blue points). The difference in linewidth, however, varies by less than 0.3\,meV, demonstrating that the thermal activation of intervalley phonon scattering remains negligible and does not alter the lifetime of the $1s\text{ KK}$ exciton polaron in MoSe$_2$.

To contrast these findings with an optically dark material, we now examine monolayer WSe$_2$. Unlike MoSe$_2$, it hosts an optically dark $\text{KK}'$ exciton ground state that is  located at about 60\,meV below the bright $\text{KK}$ exciton state \cite{Selig2018} (see inset in Fig. \ref{fig5}e). Due to this large energy offset, the $\text{KK}$ exciton can scatter into the $\text{KK}'$ exciton states by emitting a K-point phonon, provided that the energy difference between the two exciton states matches the energy of TA or LA phonons \cite{Kim2014}. As the energy of the $\text{KK}'$ exciton increases with momentum, a matching final state exists for any initial momentum of the $\text{KK}$ exciton. This means that the phonon-emission channel remains open at all momenta, which, as for the intravalley case (Fig. \ref{fig3}), we expect to substantially broaden the spectral function and shorten the lifetime of the $\text{KK}$ acoustic polaron.

To verify this prediction, and demonstrate polaron lifetime modifications induced by intervalley coupling, we calculate the spectral function $A_{\mathbf{Q}}$ for a WSe$_2$ monolayer considering the phonon-mediated coupling between the $1s\text{ KK}$ and $1s\text{ KK}'$ excitons. Figure \ref{fig5}e displays a 2D contour plot of $A_\Qv$ at $100\,\text{K}$, revealing that the $\text{KK}$ polaron peak is redshifted by $\sim4\text{ meV}$ relative to the bare $\text{KK}$ exciton dispersion (white dashed line). This redshift is about half that obtained in MoSe$_2$. In addition, the spectral functions at $Q = 0$ with (red curve in Fig. \ref{fig5}f) and without (blue curve in Fig. \ref{fig5}f) intervalley coupling are almost identical in energy, showing that intervalley coupling causes a negligible additional shift in WSe$_2$, in contrast to MoSe$_2$ (compare Figs. \ref{fig5}c and g). This difference originates from the intervalley self-energies $\Sigma_{\mu\nu}$ entering the spectral function at $Q=0$. In WSe$_2$, the opening of intervalley phonon scattering channels results in real and imaginary parts of $\Sigma_{\mu\nu}$ that are comparable in magnitude, yet both remain smaller than their counterparts in MoSe$_2$ (Supplementary Note 2). Thus in WSe$_2$ the coupling mainly broadens the resonance rather than shifting it. Extracting the FWHM of both spectra in Fig. \ref{fig5}f, we find that intervalley coupling increases the linewidth from 4.1\,meV to 5.2\,meV, i.e. by $\Delta\text{FWHM}= 1.1\,\text{meV}$, considerably more than in MoSe$_2$. We attribute this additional broadening to KK-KK' scattering via emission of K-point acoustic phonons, which limits the polaron lifetime to about 0.55\,ps. More importantly, $\Delta\text{FWHM}$ remains nearly constant at $\sim 1.1\,\text{meV}$ over the entire temperature range (see Fig. \ref{fig5}h), indicating that intervalley scattering remains active even at $10\,\text{K}$, which is characteristic for a phonon emission channel. These results suggest that exciton polarons in a WSe$_2$ monolayers possess a reduced lifetimes compared to those in MoSe$_2$ owing to intervalley phonon emission, thus limiting their stability as well-defined quasiparticles.

\section*{\label{discussion}  Discussion}

In summary, we have introduced a theoretical framework to calculate exciton-polaron properties in monolayer transition metal dichalcogenides. Applying this framework to $1s$ excitons interacting with acoustic phonons in MoSe$_2$ and WSe$_2$ monolayers, we obtained the dispersion relation and lifetime of the resulting exciton polarons. These calculations not only capture the expected polaron energy redshift, reduced group velocity and enhanced effective mass, but also reveal a novel feature in the exciton spectral function that separates two distinct momentum regimes, one characterized by bound exciton polarons and the other dominated by exciton–phonon scattering. By extending the method to incorporate intervalley excitons, we reveal that $\text{KK}$ acoustic exciton polarons in monolayer WSe$_2$ exhibit reduced lifetimes relative to their MoSe$_2$ counterparts. This stems from the efficient scattering of $\text{KK}$ excitons into the lower-lying $\text{ KK}'$ exciton states via emission of K-point acoustic phonons. Thus, our findings establish intervalley scattering as an important mechanism determining exciton-polaron dynamics in TMDs, and demonstrate that the underlying exciton energy landscape is crucial for understanding polaron phenomena. The developed theoretical framework is broadly generalizable e.g. to exciton polarons in 2D perovskites \cite{Paulina2024, Kandada2020}, or moiré superlattices \cite{Knorr2024} and establishes a robust method for investigating exciton polarons in technologically promising two-dimensional materials.

\section*{\label{methods} Methods}

\subsection*{ Wannier equation} 

To determine the exciton energies $\Ex_{\mu\Qv}$ and wavefunctions $\Psi_{\mu \kv}$, we apply the effective mass approximation, assuming parabolic band dispersions at the high-symmetry extrema of the monolayer Brillouin zone \cite{Kira2006}. Within this framework, we solve the Wannier equation \cite{Malic2014}
\begin{equation}
\label{Wannier}
\frac{\hbar^2 k^2}{2m^{\text{r}}_{\mu}} \Psi_{\mu \kv} - \sum_{\kv'} V_{\kv'-\kv}\Psi_{\mu \kv'} = E^{\text{b}}_{\mu} \Psi_{\mu \kv}.
\end{equation}
Here, $\mu = (n^{\mu}, \xi^{\mu}_e, \xi^{\mu}_h)$ is a composite index, where $n^{\mu}$ denotes the principal quantum number of the exciton Rydberg series, and $\xi^{\mu}_e$ ($\xi^{\mu}_h$) specifies the high-symmetry valley for the electron (hole). In Eq. \eqref{Wannier}, $\kv$ represents the relative electron–hole momentum, $(m^{\text{r}}_\mu)^{-1} = (m^e_\mu)^{-1} + (m^h_\mu)^{-1}$ is the inverse of the reduced exciton mass, and $m^e_\mu$ ($m^h_\mu$) is the electron (hole) mass. To account for the dielectric screening and finite thickness of the TMD monolayer, we describe the electron–hole Coulomb potential $V_{\kv'-\kv}$ using a Keldysh-like model \cite{Keldysh1979}. The material-specific parameters used to solve Eq. \eqref{Wannier} were obtained from DFT calculations \cite{Kormanyos2015} and are provided in Supplementary Note 2.
Once the exciton binding energies $E^{\text{b}}_\mu$ are obtained by solving the Wannier equation (Eq. \eqref{Wannier}), the exciton energy dispersion relations are determined as $\Ex_{\mu\Qv} = \hbar^2 Q^2 / \left(2 \mx_\mu \right) + E^{\text{g}}_\mu + E^{\text{b}}_\mu$, where $\mx_\mu = m^e_\mu + m^h_\mu$ is the mass of the exciton and $E^{\text{g}}_\mu$ is the energy gap between the valence and conduction bands.

\subsection*{Matsubara Green's function for excitons and Dyson equation}
To derive the secular equation shown in the main text (Eq. \eqref{Dyson}), we employ the many-body Green's function together with the equation-of-motion formalisms \cite{Walecka2003,Bruus2004,Nolting2009}. Specifically, we introduce the Matsubara Green's function for excitons as
\begin{align}
\label{Green1}
\Gm{\mu\Qv}{\nu\Qv}(\tau,\tau') &= - \frac{i}{\hbar} \average{\hat{T}_\tau \destruye{X}_{\mu\Qv}(\tau) \crea{X}_{\nu\Qv}(\tau')},
\end{align}
where $\hat{T}_\tau$ denotes the imaginary-time-ordering operator \cite{Nolting2009} and $\langle \hat{O} \rangle $ represents the expectation value of operator $\hat{O}$. Here, $\destruye{X}_{\mu\Qv}(\tau)$ and $\crea{X}_{\nu\Qv}(\tau)$ are the exciton annihilation and creation operators in the Heisenberg picture, respectively. These operators describe the annihilation of an exciton in the state $\mu$ with center-of-mass momentum $\Qv$ at time $\tau$, and the creation of an exciton in the state $\nu$ with the same momentum at time $\tau'$. 

Solving the equations of motion of $\Gm{\mu\Qv}{\nu\Qv}(\tau,\tau')$ in the frequency domain and performing the standard analytic continuation from Matsubara to real frequencies, one finds the following Dyson equation for the retarded Green's function $\Gr{\mu\Qv}{\nu\Qv} (E)$:
\begin{align}
\label{Green2}
\sum_{\lambda} \left[ (E - \Ex_{\mu\Qv})\delta_{\mu\lambda} - \Sigma_{\mu\lambda\Qv}(E) \right] \Gr{\lambda\Qv}{\nu\Qv} (E) = \delta_{\mu \nu},
\end{align}
where $\lambda$ labels exciton states and $\Sigma_{\mu\nu\Qv}(E)$ is the Fan-Migdal self-energy
\begin{align}
\label{Green3}
&\Sigma_{\mu\nu\Qv}(E) =  \sum_{\xi, \alpha \qv} \mathcal{D}_{\alpha \qv}^{{\xi \mu}} \mathcal{D}_{\alpha, -\qv}^{{\nu \xi}} \times \\
\nonumber
&\times \left( \frac{ \averageo{\hat{n}_{\alpha\qv}}}{E - \Ex_{\xi \Qv-\qv}+ \hbar \Omega_{\alpha \qv} + i \eta}+ \right. \\
\nonumber
& \left. + \frac{1+\averageo{\hat{n}_{\alpha\qv}}}{E - \Ex_{\xi \Qv-\qv} - \hbar \Omega_{\alpha \qv} + i \eta} \right).
\end{align}
Here, $\averageo{\hat{n}_{\alpha\qv}}$ is the Bose-Einstein distribution for the occupation of phonons, $\hbar \Omega_{\alpha \qv}$ is the energy of the $\alpha$ phonon at momentum $\qv$ and $\eta$ is a numerical broadening parameter introduced to satisfy causality. The poles of $\Gr{\lambda\Qv}{\nu\Qv}$ correspond to the polaron energies $\Ep_\Qv$, which are obtained by requiring the determinant of the bracketed expression in Eq. \eqref{Green2} to vanish, yielding Eq. \eqref{Dyson}. In Supplementary Notes 1 and 2, we provide further details on the derivation of Eqs. \eqref{Green2}-\eqref{Green3}, as well as the material-specific parameters and methods used in the numerical simulations of each figure shown in the main text.

\begin{acknowledgments}
C. M.-E. acknowledges the Alexander von Humboldt Foundation for financial support through a Humboldt Research Fellowship. Furthermore, we acknowledge funding from the Deutsche Forschungsgemeinschaft (DFG) via the regular project 535253440 and by the State of Hesse via the LOEWE Exploration Programme.
\end{acknowledgments}

\section*{\label{references} References}
\end{document}


\maketitle

\section*{Supplementary Note 1. Derivation of Dyson equation and the Fan-Migdal self-energy for excitons}\label{note1}

In this Supplementary Note, we derive Eqs. (2) and (7) presented in the main text combining the equation-of-motion and many-body Green's function formalisms \cite{Walecka2003,Bruus2004,Nolting2009}. We begin by introducing the exciton Matsubara Green's function as
\begin{equation}
\label{s1}
\Gm{\mu\Qv}{\nu\Qv'}(\tau,\tau') = - \frac{1}{\hbar} \average{\hat{T}_\tau \destruye{X}_{\mu\Qv}(\tau) \crea{X}_{\nu\Qv'}(\tau')},
\end{equation}
where $\hat{T}_\tau$ denotes the imaginary time-ordering operator \cite{Nolting2009}, and $\tau$, $\tau'$ are real variables parametrizing imaginary times via $t=-i\tau$, $t'=-i\tau'$. The operators $\destruye{X}_{\mu\Qv}(\tau)$ and $\crea{X}_{\nu\Qv'}(\tau')$ are exciton annihilation and creation operators in the Heisenberg picture, and the labels $\mu\Qv$ and $\nu\Qv'$ denote excitonic states with center-of-mass momenta $\hbar \Qv$ and $\hbar \Qv'$, respectively. The expectation value $\average{\hat{T}_\tau \destruye{X}_{\mu\Qv}(\tau) \crea{X}_{\nu\Qv'}(\tau')}$ is calculated via the canonical partition function $Z = \Tr{e^{-\beta_T \Hxp}}$ as
\begin{equation}
\label{s2}
\average{\hat{T}_\tau \destruye{X}_{\mu\Qv}(\tau) \crea{X}_{\nu\Qv'}(\tau')}= \frac{\Tr{e^{-\beta_T \Hxp} \hat{T}_\tau \destruye{X}_{\mu\Qv}(\tau) \crea{X}_{\nu\Qv'}(\tau')}}{\Tr{e^{-\beta_T \Hxp}}},
\end{equation}
with $\beta_T = 1/(k_{\text{B}}T)$, $T$ the temperature of the system and $\Hxp$ is the exciton-phonon Hamiltonian determined by Eq. (1) in the main text:
\begin{equation}
\label{s3}
\Hxp = \sum_{\mu, \Qv} \Ex_{\mu \Qv}\crea{X}_{\mu \Qv} \destruye{X}_{\mu \Qv} + \sum_{\alpha, \kv} \hbar \Omega_{\alpha \kv}\,  \crea{b}_{\alpha \kv} \destruye{b}_{\alpha \kv}
+\sum_{\mu \nu \Qv, \alpha \qv} \mathcal{D}_{\alpha \qv}^{{\nu \mu}} \crea{X}_{\mu \Qv + \qv} \destruye{X}_{\nu \Qv} (\destruye{b}_{\alpha \qv} + \crea{b}_{\alpha, -\qv}).
\end{equation}
In this Hamiltonian, all operators are defined within the Schr\"odinger picture. The operators $\crea{b}_{\alpha \qv}$ and $\destruye{b}_{\alpha \qv}$ are the corresponding bosonic creation and annihilation operators for a phonon in the acoustic mode $\alpha$ with momentum $\hbar \qv$. The terms $\Ex_{\mu \vect{Q}}$ and $\hbar \Omega_{\alpha \qv}$ represent the exciton and phonon energy dispersion relations, respectively. The exciton dispersion is obtained from the Wannier equation (see Methods in the main text), while the phonon dispersion relation and the exciton–phonon matrix elements $\mathcal{D}_{\alpha \mathbf{q}}^{\nu \mu}$ are detailed in Supplementary Note 2.

Owing to the time-independence and spatial translational invariance of $\Hxp$, the Green's function (Eq. \eqref{s1}) is invariant under imaginary-time translations and spatial translations. Consequently, it is diagonal in the center-of-mass momentum and depends only on the imaginary-time difference as $\Gm{\mu\Qv}{\nu\Qv'}(\tau,\tau') = \delta_{\Qv\Qv'} \Gm{\mu\Qv}{\nu\Qv}(\tau-\tau',0)$. Thus, one can redefine the exciton Matsubara Green's function from Eq. \eqref{s1} to 
\begin{equation}
\label{s4}
\Gm{\mu\Qv}{\nu\Qv}(\tau,\tau') = \Gm{\mu\Qv}{\nu\Qv}(\tau-\tau',0) = - \frac{1}{\hbar} \average{\hat{T}_\tau \destruye{X}_{\mu\Qv}(\tau-\tau') \crea{X}_{\nu\Qv}(0)}.
\end{equation}

Differentiating Eq. \eqref{s4} with respect to $\tau$ and applying the Heisenberg equation of motion to determine the time evolution of an arbitrary operator $\hat{O}(\tau)$,
\begin{equation}
\label{s5}
\frac{\partial \hat{O}(\tau)}{\partial \tau} = -\frac{1}{\hbar} [\hat{O}(\tau), \Hxp],
\end{equation}
yields the following equation of motion for $\Gm{\mu\Qv}{\nu\Qv}(\tau,\tau')$
\begin{equation}
\label{s6}
-\left(\hbar \frac{\partial}{\partial \tau} + \Ex_{\mu \Qv} \right)  \Gm{\mu\Qv}{\nu\Qv}(\tau,\tau') = \delta(\tau-\tau')\delta_{\mu \nu} + \sum_{\xi, \alpha \qv} \mathcal{D}_{\alpha \qv}^{{\xi \mu}} F_{\xi\Qv-\qv,\alpha\qv,\nu\Qv} (\tau,\tau'),
\end{equation}
where $\delta(\tau-\tau')$ represents the Dirac delta function, $\delta_{\mu \nu}$ represents the delta Kronecker delta function, $\xi$ labels exciton states and  
\begin{equation}
\label{s7}
F_{\xi\Qv-\qv,\alpha\qv,\nu\Qv} (\tau,\tau')= -\frac{1}{\hbar} \average{\hat{T
}_\tau \destruye{X}_{\xi \Qv - \qv}(\tau) \hat{B}_{\alpha \qv}(\tau) \crea{X}_{\nu \Qv}(\tau')},
\end{equation}
with $\hat{B}_{\alpha \qv} = \destruye{b}_{\alpha\qv} + \crea{b}_{\alpha,-\qv}$ the phonon displacement operator. Equation \eqref{s7} takes the same mathematical form as Eq. \eqref{s4} and can be identified as the exciton-phonon Green's function (or propagator). As demonstrated below, solving the system of coupled equations \eqref{s6}–\eqref{s7} for $\Gm{\mu\Qv}{\nu\Qv}$ and $F_{\xi\Qv-\qv,\alpha\qv,\nu\Qv}$ yields a Dyson-like equation.

To solve the equation of motion \eqref{s6}, we first derive an analytical expression for Eq. \eqref{s7} by separating $\hat{H}_{\text{xp}}$ into non-interacting and interaction terms,
\begin{equation}
\label{s8}
\Hxp = \Ho + \Hint,
\end{equation}
with 
\begin{align}
\label{s9}
\Ho = \sum_{\mu,\Qv} \Ex_{\mu \Qv}\crea{X}_{\mu \Qv} \destruye{X}_{\mu \Qv} + \sum_{\alpha, \kv} \hbar \Omega_{\alpha \kv}\,  \crea{b}_{\alpha \kv} \destruye{b}_{\alpha \kv}, \\
\label{s10}
\Hint = \sum_{\mu \nu \Qv, \alpha \qv} \mathcal{D}_{\alpha \qv}^{{\nu \mu}} \crea{X}_{\mu \Qv + \qv} \destruye{X}_{\nu \Qv} (\destruye{b}_{\alpha \qv} + \crea{b}_{\alpha, -\qv}).
\end{align}
We then transform exciton and phonon operators in Eq. \eqref{s7} to the interaction picture via the transformation $\hat{O}_{\text{I}}(\tau) = e^{\frac{\tau}{\hbar} \hat{H}_0} \hat{O} e^{-\frac{\tau}{\hbar} \hat{H}_0}$. Transforming the operators to the interaction picture allows us to compute the expectation value in Eq. \eqref{s7} using the Gell-Mann and Low theorem \cite{GellMann1951,Walecka2003} as:
\begin{align}
\label{s11}
\average{\hat{T
}_\tau \destruye{X}_{\xi \Qv - \qv}(\tau) \hat{B}_{\alpha \qv}(\tau) \crea{X}_{\nu \Qv}(\tau')}
&= \frac{\Tr{e^{-\beta_T \Ho} \hat{T}_\tau e^{-\frac{1}{\hbar} \int_{0}^{\hbar \beta_T} \text{d}\tau_1 \HintI(\tau_1)} \destruye{X}_{\xi \Qv - \qv, \text{I}}(\tau) \hat{B}_{\alpha \qv, \text{I}}(\tau) \crea{X}_{\nu \Qv, \text{I}}(\tau')}}{\Tr{e^{-\beta_T \Ho} \hat{T} e^{-\frac{1}{\hbar} \int_{0}^{\hbar \beta_T} \text{d}\tau_1 \HintI(\tau_1)}}}\\
\nonumber
&= \frac{ \averageo{\hat{T} e^{-\frac{1}{\hbar} \int_{0}^{\hbar \beta_T} \text{d}\tau_1 \HintI(\tau_1)} \destruye{X}_{\xi \Qv - \qv, \text{I}}(\tau) \hat{B}_{\alpha \qv, \text{I}}(\tau) \crea{X}_{\nu \Qv, \text{I}}(\tau')}}{\averageo{\hat{T} e^{-\frac{1}{\hbar} \int_{0}^{\hbar \beta_T} \text{d}\tau_1 \HintI(\tau_1)} }}.
\end{align}
Here, the subindex ``I'' indicates operators defined in the interaction picture, while the subindex ``0'' in the expectation value $\averageo{\cdots}$ denotes a thermal average evaluated with respect to the non-interacting Hamiltonian $\Ho$. We note that exciton Matsubara Green's function, as defined in Eq. \eqref{s2}, is periodic with period $\hbar \beta_T$ over the interval $[0,\hbar \beta_T]$.

The solution to Eq. \eqref{s11} is an infinite series in which each term represents a distinct exciton–phonon interaction. These interactions are typically represented by connected Feynman diagrams, as detailed in Refs. \cite{Walecka2003,Bruus2004}. As mentioned in the main text, we restrict our analysis to interactions where an exciton emits and subsequently reabsorbs a single phonon. A mathematical expression for these interactions can be derived by expanding the exponential $e^{-\frac{1}{\hbar} \int_{0}^{\hbar \beta_T} \text{d}\tau_1 \HintI(\tau_1)}$ to first order:
\begin{equation}
\label{s12}
e^{-\frac{1}{\hbar} \int_{0}^{\hbar \beta_T} \text{d}\tau_1 \HintI(\tau_1)} \approx 1 -\frac{1}{\hbar} \int_{0}^{\hbar \beta_T} \text{d}\tau_1 \HintI(\tau_1),
\end{equation}
and subsequently inserting this expansion into Eq. \eqref{s11}:
\begin{align}
\label{s13}
&\average{\hat{T}_\tau \destruye{X}_{\xi \Qv - \qv}(\tau) \hat{B}_{\alpha \qv}(\tau) \crea{X}_{\nu \Qv}(\tau')} \approx \\
\nonumber
&\approx \frac{\averageo{\hat{T}_\tau\destruye{X}_{\xi \Qv - \qv, \text{I}}(\tau) \hat{B}_{\alpha \qv, \text{I}}(\tau) \crea{X}_{\nu \Qv, \text{I}}(\tau')} - \frac{1}{\hbar} \averageo{ \hat{T}_\tau \int_{0}^{\hbar\beta_T} \text{d}\tau_1 \HintI(\tau_1) \destruye{X}_{\xi \Qv - \qv, \text{I}}(\tau) \hat{B}_{\alpha \qv, \text{I}}(\tau) \crea{X}_{\nu \Qv, \text{I}}(\tau') }}{\averageo{\hat{T}_\tau e^{-\frac{1}{\hbar} \int_{0}^{\hbar \beta_T} \text{d}\tau_1 \HintI(\tau_1)} }}.
\end{align}
Applying the generalized (finite-temperature) Wick's theorem \cite{Matsubara1955,Walecka2003} and neglecting coherent phonon generation (such that $\averageo{\hat{B}_{\alpha\qv,\text{I}}} = 0$), one can show that the first term in the numerator of Eq. \eqref{s13} is zero, that is, $\averageo{\hat{T}_\tau\destruye{X}_{\xi \Qv - \qv, \text{I}}(\tau) \hat{B}_{\alpha \qv, \text{I}}(\tau) \crea{X}_{\nu \Qv, \text{I}}(\tau')} = 0$. Thus, it is sufficient to calculate Eq. \eqref{s13} as:
\begin{equation}
\label{s14}
\average{\hat{T}_\tau \destruye{X}_{\xi \Qv - \qv}(\tau) \hat{B}_{\alpha \qv}(\tau) \crea{X}_{\nu \Qv}(\tau')} \approx 
- \frac{1}{\hbar} \frac{ \averageo{ \hat{T}_\tau \int_{0}^{\hbar\beta_T} \text{d}\tau_1 \HintI(\tau_1) \destruye{X}_{\xi \Qv - \qv, \text{I}}(\tau) \hat{B}_{\alpha \qv, \text{I}}(\tau) \crea{X}_{\nu \Qv, \text{I}}(\tau') }}{\averageo{\hat{T}_\tau e^{-\frac{1}{\hbar} \int_{0}^{\hbar \beta_T} \text{d}\tau_1 \HintI(\tau_1)} }}.
\end{equation}
Furthermore, transforming $\Hint$ (Eq. \eqref{s10}) to the interaction picture and substituting it into the numerator of Eq. \eqref{s14} yields the following expression:
\begin{align}
\label{s15}
&\int_{0}^{\hbar\beta_T} \text{d}\tau_1  \averageo{ \hat{T}_\tau \HintI(\tau_1) \destruye{X}_{\xi \Qv - \qv, \text{I}}(\tau) \hat{B}_{\alpha \qv, \text{I}}(\tau) \crea{X}_{\nu \Qv, \text{I}}(\tau') }=  \\
%
\nonumber
&=\sum_{\gamma \zeta \vect{Q}'', \alpha' \kv''} \mathcal{D}_{\alpha' \kv''}^{{\zeta \gamma}} \int_{0}^{\hbar \beta_T} \text{d}\tau_1\, \averageo{ \hat{T}_\tau \crea{X}_{\gamma \vect{Q}''+ \kv'', \text{I}}(\tau_1) \destruye{X}_{\zeta \vect{Q}'', \text{I}}(\tau_1) \destruye{B}_{\alpha' \kv'', \text{I}}(\tau_1) \destruye{X}_{\xi \Qv - \qv, \text{I}}(\tau) \hat{B}_{\alpha \qv, \text{I}}(\tau) \crea{X}_{\nu \Qv, \text{I}}(\tau')  },
\end{align}
where $\gamma$ and $\zeta$ label exciton states, and $\alpha'$ phonon modes. Plugging Eq. \eqref{s15} into Eq. \eqref{s14}, then applying the generalized Wick's theorem and considering the terms that yields connected Feynman diagrams, one finds the following expression:
\begin{align}
\label{s16}
&\average{\hat{T}_\tau \destruye{X}_{\xi \Qv - \qv}(\tau) \hat{B}_{\alpha \qv}(\tau) \crea{X}_{\nu \Qv}(\tau')} \approx - \frac{1}{\hbar} \sum_{\gamma \zeta \vect{Q}'', \alpha' \kv''} \mathcal{D}_{\alpha' \kv''}^{{\zeta \gamma}} \int_{0}^{\hbar \beta_T} \text{d}\tau_1\, \times \\
\nonumber
&
\times \left\{\averageo{ \hat{T}_\tau \crea{X}_{\gamma \vect{Q}''+ \kv'', \text{I}}(\tau_1) \destruye{X}_{\zeta \vect{Q}'', \text{I}}(\tau_1)}\averageo{\hat{T}_\tau \destruye{B}_{\alpha' \kv'', \text{I}}(\tau_1)  \hat{B}_{\alpha \qv, \text{I}}(\tau)  } \averageo{\hat{T}_\tau \destruye{X}_{\xi \Qv - \qv, \text{I}}(\tau) \crea{X}_{\nu \Qv, \text{I}}(\tau')  } \right.\\
\nonumber
&\left. +\averageo{ \hat{T}_\tau \crea{X}_{\gamma \vect{Q}''+ \kv'', \text{I}}(\tau_1) \destruye{X}_{\xi \Qv - \qv, \text{I}}(\tau)} \averageo{\hat{T}_\tau \destruye{B}_{\alpha' \kv'', \text{I}}(\tau_1)  \hat{B}_{\alpha \qv, \text{I}}(\tau)  } \averageo{\hat{T}_\tau \destruye{X}_{\zeta \vect{Q}'', \text{I}}(\tau_1)\crea{X}_{\nu \Qv, \text{I}}(\tau') } \right\}.
\end{align}
To derive this expression, crossover terms such as $\averageo{\hat{X}\hat{B}}$ are neglected because they factorize into $\averageo{\hat{X}}\averageo{\hat{B}}$, and we assume no coherent phonon generation, i.e., $\averageo{\hat{B}} = 0$. Furthermore, the term in the second line of Eq. \eqref{s16} describes non-interacting, instantaneous exciton transitions. As our focus is strictly on exciton–phonon interactions via phonon absorption and emission, such terms are neglected. Thus, Eq. \eqref{s16} simplifies to:
\begin{align}
\label{s17}
&\average{\hat{T}_\tau \destruye{X}_{\xi \Qv - \qv}(\tau) \hat{B}_{\alpha \qv}(\tau) \crea{X}_{\nu \Qv}(\tau')} \approx - \frac{1}{\hbar} \sum_{\gamma \zeta \vect{Q}'', \alpha' \kv''} \mathcal{D}_{\alpha' \kv''}^{{\zeta \gamma}} \int_{0}^{\hbar \beta_T} \text{d}\tau_1\, \times \\
\nonumber
&\times \left\{\averageo{ \hat{T}_\tau \crea{X}_{\gamma \vect{Q}''+ \kv'', \text{I}}(\tau_1) \destruye{X}_{\xi \Qv - \qv, \text{I}}(\tau)} \averageo{\hat{T}_\tau \destruye{B}_{\alpha' \kv'', \text{I}}(\tau_1)  \hat{B}_{\alpha \qv, \text{I}}(\tau)  } \averageo{\hat{T}_\tau \destruye{X}_{\zeta \vect{Q}'', \text{I}}(\tau_1)\crea{X}_{\nu \Qv, \text{I}}(\tau') } \right\}.
\end{align}
A further simplification is achieved by introducing the Matsubara propagators for free excitons and phonons
\begin{equation}
\label{s18}
\Gm{\mu\Qv}{\nu\Qv'}^0(\tau,\tau') = \Gm{\mu\Qv}{\nu\Qv'}^0(\tau-\tau',0) = - \frac{1}{\hbar} \averageo{\hat{T}_\tau \destruye{X}_{\mu\Qv}(\tau) \crea{X}_{\nu\Qv'}(\tau')} \, \delta_{\mu\nu} \, \delta_{\Qv\Qv'},
\end{equation}
and
\begin{equation}
\label{s19}
D^0_{\beta\kv,\alpha\kv'}(\tau_1,\tau_2) = D^0_{\beta\kv,\alpha\kv'}(\tau_1-\tau_2,0) =  -\frac{1}{\hbar} \averageo{ \hat{T}_\tau \hat{B}_{\beta \kv}(\tau_1) \destruye{B}_{\alpha \kv'}(\tau_2)} \, \delta_{\beta\alpha} \, \delta_{\kv,-\kv'},
\end{equation}
respectively. Note that these propagators are diagonal in momentum and depend only on the time difference due to the translational invariance and time independence of the non-interacting Hamiltonian $\Ho$ (Eq. \eqref{s9}). We thus focus our derivation on these diagonal elements. In the low-density limit, excitons creation and annihilation operators obey standard bosonic commutation relations, allowing us to reorder the operators within the thermal averages of Eq. \eqref{s17} and match the definitions from Eqs. \eqref{s18} and \eqref{s19}. Thus, expressing Eq. \eqref{s17} in terms of the free exciton and phonon propagators and inserting the expression into Eq. \eqref{s7}, one can derive the following expression for the exciton–phonon propagator $F_{\xi\Qv-\qv,\alpha\qv,\nu\Qv}$ (Eq. \eqref{s7}):
\begin{align}
\label{s20}
&F_{\xi\Qv-\qv,\alpha\qv,\nu\Qv} (\tau,\tau') = -\frac{1}{\hbar}\average{\hat{T}_\tau \destruye{X}_{\xi \Qv - \qv}(\tau) \hat{B}_{\alpha \qv}(\tau) \crea{X}_{\nu \Qv}(\tau')} \approx \\
\nonumber
& \approx -\hbar \mathcal{D}_{\alpha, -\qv}^{{\nu \xi}} \int_{0}^{\hbar \beta_T} \text{d}\tau_1\,  \Gm{\xi \Qv-\qv}{\xi \Qv-\qv}^0(\tau,\tau_1) D^0_{\alpha \qv,\alpha, -\qv}(\tau,\tau_1) \Gm{\nu \Qv}{\nu \Qv}^0(\tau_1,\tau'). 
\end{align}

Using this expression, one can deduce the following equation of motion for $\Gm{\mu\Qv}{\nu\Qv}(\tau,\tau')$ (Eq. \eqref{s6})
\begin{align}
\label{s21}
&-\left(\hbar \frac{\partial}{\partial \tau} + \Ex_{\mu \Qv} \right)  \Gm{\mu\Qv}{\nu\Qv}(\tau,\tau') \approx \delta(\tau-\tau')\delta_{\mu \nu} - \\
%
\nonumber
& -\hbar \sum_{\xi, \alpha \qv} \mathcal{D}_{\alpha \qv}^{{\xi \mu}} \mathcal{D}_{\alpha, -\qv}^{{\nu \xi}} \int_{0}^{\hbar \beta_T} \text{d}\tau_1\,  \Gm{\xi \Qv-\qv}{\xi \Qv-\qv}^0(\tau,\tau_1) D^0_{\alpha \qv,\alpha, -\qv}(\tau,\tau_1) \Gm{\nu\Qv}{\nu\Qv}^0(\tau_1,\tau'),
\end{align}
which can be solved by exploiting the periodicity of Matsubara Green's functions with period $\hbar \beta_T$ over the interval $[0, \hbar \beta_T]$. Specifically, we expand $\Gm{\mu\Qv}{\nu\Qv}, \Gm{\mu\Qv}{\nu\Qv}^0$, $D^0_{\alpha \qv,\alpha, -\qv}$ and $\delta(\tau-\tau')$ into their Fourier representations:
\begin{align}
\label{s22}
&\Gm{\mu\Qv}{\nu\Qv}(\tau, \tau') = \frac{1}{\hbar \beta_T} \sum_{n=-\infty}^{\infty}  g_{\mu\Qv,\nu\Qv}(E_n) e^{-\frac{i}{\hbar}E_n(\tau-\tau')},  \\
\label{s23}
&\Gm{\mu\Qv}{\nu\Qv}^0(\tau, \tau') = \frac{1}{\hbar \beta_T} \sum_{n=-\infty}^{\infty} g^0_{\mu\Qv,\nu\Qv}(E_n)e^{-\frac{i}{\hbar}E_n(\tau-\tau')}, \\
\label{s24}
&D^0_{\alpha \qv,\alpha, -\qv}(\tau, \tau') = \frac{1}{\hbar \beta_T}\sum_{n=-\infty}^{\infty} d^0_{\alpha \qv,\alpha, -\qv}(E_n) e^{-\frac{i}{\hbar}E_n(\tau-\tau')}, \\
\label{s25}
&\delta(\tau-\tau') = \frac{1}{\hbar \beta_T} \sum_{n=-\infty}^{\infty} e^{-\frac{i}{\hbar} E_n(\tau-\tau')},
\end{align}
where $E_n = \hbar\omega_n =  2\pi n /\beta_T$, $\omega_n$ is the conventional Matsubara frequency for bosons, $n\in\mathbb{Z}$, and the \textit{n}th Fourier coefficients are defined as
\begin{align}
\label{s26}
&g_{\mu\Qv,\nu\Qv}(E_n) =\int_0^{\hbar \beta_T} \text{d}\tau \Gm{\mu\Qv}{\nu\Qv}(\tau,0) e^{\frac{i}{\hbar} E_n \tau},  \\
\label{s27}
&g^0_{\mu\Qv,\nu\Qv}(E_n) =\int_0^{\hbar \beta_T} \text{d}\tau \Gm{\mu\Qv}{\nu\Qv}^0(\tau,0) e^{\frac{i}{\hbar} E_n \tau}, \\
\label{s28}
&d^0_{\alpha \qv,\alpha, -\qv}(E_n) =\int_0^{\hbar \beta_T} \text{d}\tau D^0_{\alpha \qv,\alpha, -\qv}(\tau,0) e^{\frac{i}{\hbar} E_n \tau}.
\end{align}
Substituting the Fourier expansions (Eqs. \eqref{s22}-\eqref{s25}) into Eq. \eqref{s21}, rearranging terms and using the relation $\int_{0}^{\hbar\beta_T} \text{d}\tau \, e^{\frac{i}{\hbar} \tau (E_n-E_m)}=\delta_{mn} \hbar\beta_T$, together with the orthogonality of the Fourier basis vectors, one can deduce the following relation for $g_{\mu\Qv,\nu\Qv}$
\begin{align}
\label{s29}
&\left(i E_n  - \Ex_{\mu \Qv} \right)  g_{\mu\Qv,\nu\Qv}(E_n) \approx \delta_{\mu \nu} - \sum_{\xi, \alpha \qv} \mathcal{D}_{\alpha \qv}^{{\xi \mu}} \mathcal{D}_{\alpha, -\qv}^{{\nu \xi}}\,  g_{\nu\Qv, \nu\Qv}^0(E_n) \times\\
\nonumber
&\times \left( \frac{1}{\beta_T}\sum_{m=-\infty}^{\infty}  g_{\xi\Qv-\qv,\xi\Qv-\qv}^0(E_m) \, d^0_{\alpha \qv,\alpha, -\qv}(E_n -E_m) \right). 
\end{align}
\sloppy We have thus transformed the problem of finding $G_{\mu\Qv,\nu\Qv}$ from solving a differential equation (Eq. \eqref{s21}) to solving an algebraic equation (Eq. \eqref{s29}) for $g_{\mu\Qv,\nu\Qv}$. Furthermore, using the Baker-Campbell-Hausdorff formula (Hadamard's lemma), one can show that phonon creation and annihilation operators in the interaction picture take the following form: $\crea{b}_{\alpha\qv, \text{I}} (\tau)= e^{\frac{1}{\hbar} \Ho \tau} \crea{b}_{\alpha\qv} e^{ - \frac{1}{\hbar} \Ho \tau}  = \crea{b}_{\alpha\qv} e^{\Omega_{\alpha \qv} \tau}$ and $\destruye{b}_{\alpha\qv, \text{I}} (\tau)= e^{\frac{1}{\hbar} \Ho \tau} \destruye{b}_{\alpha\qv} e^{ - \frac{1}{\hbar} \Ho \tau}  = \destruye{b}_{\alpha\qv} e^{- \Omega_{\alpha \qv} \tau}$. Substituting these expressions into Eq. \eqref{s19} and applying the definition of the imaginary time-ordering operator leads to the following expression for the free phonon propagator $D^0_{\alpha\qv,\alpha,-\qv}$
\begin{align}
\label{s30}
-\hbar D^0_{\alpha\qv,\alpha,-\qv}(\tau_1,\tau_2) &= \Theta(\tau_1-\tau_2) \left[  (1 + \averageo{\hat{n}_{\alpha\qv}} ) e^{- \Omega_{\alpha\qv} (\tau_1-\tau_2)}  + \averageo{\hat{n}_{\alpha \qv}} e^{\Omega_{\alpha\qv} (\tau_1 -\tau_2)}  \right] + \\
%
\nonumber
& + \Theta(\tau_2 - \tau_1) \left[ \averageo{\hat{n}_{\alpha\qv}}  e^{\Omega_{\alpha\qv} (\tau_2-\tau_1)} + (1 + \averageo{\hat{n}_{\alpha\qv}}) e^{-\Omega_{\alpha\qv} (\tau_2 -\tau_1)} \right],
\end{align}
where $\Theta$ is the step function and $\averageo{\hat{n}_{\alpha\qv}} =  1/ (e^{\beta_T \hbar \Omega_{\alpha \qv}} - 1)$ is the Bose-Einstein distribution for the occupation of phonons. Using Eq. \eqref{s30} to evaluate the integral in Eq. \eqref{s28} yields the closed form for the $n$th Fourier coefficient of the free phonon propagator:
\begin{align}
\label{s31}
d^{0}_{\alpha\qv,\alpha,-\qv}(E_n - E_m)=\frac{1}{i (E_n - E_m) - \hbar\Omega_{\alpha \qv}} - \frac{1}{i (E_n - E_m) + \hbar \Omega_{\alpha \qv}}.
\end{align}
Similarly, using Eqs. \eqref{s18} and \eqref{s27}, one can show that
\begin{equation}
\label{s32}
g^0_{\xi \Qv-\qv,\xi \Qv-\qv}(E_m) = \frac{1}{i E_m - \Ex_{\xi \Qv-\qv}}.
\end{equation}
As a final step to derive a solution of $g_{\mu\Qv,\nu\Qv}$, we evaluate the Matsubara sum in the second line of Eq. \eqref{s29} by substituting Eqs. \eqref{s31}-\eqref{s32}, and convert the discrete summation over Matsubara frequencies into a contour integral in the complex frequency plane, following standard techniques \cite{Bruus2004}. Then applying the residue theorem, we obtain:
\begin{align}
\label{s33}
&\frac{1}{\beta_T}\sum_{m=-\infty}^{\infty}  g_{\xi\Qv-\qv,\xi\Qv-\qv}^0(E_m) \, d^0_{\alpha \qv,\alpha, -\qv}(E_n - E_m) = \\
\nonumber
& =-\left[ \frac{ \averageo{\hat{n}_{\alpha \qv}}}{iE_n - \Ex_{\xi \Qv-\qv}+ \hbar \Omega_{\alpha \qv}} + \frac{1+\averageo{\hat{n}_{\alpha \qv}}}{iE_n - \Ex_{\xi \Qv-\qv} - \hbar \Omega_{\alpha \qv}}\right].
\end{align}
Inserting this result into Eq. \eqref{s29}, and using Eq. \eqref{s32}, yields the following relation
\begin{align}
\label{s34}
& g_{\mu\Qv,\nu\Qv}(E_n) \approx \delta_{\mu \nu} g_{\mu\Qv, \mu\Qv}^0(E_n) + g_{\mu\Qv, \mu\Qv}^0(E_n) \left[\Sigma_{\mu\nu\Qv}(E_n)\right] g_{\nu\Qv, \nu\Qv}^0(E_n),
\end{align}
where $\Sigma_{\mu\nu\mathbf{Q}}(E_n)$ is defined as
\begin{equation}
\label{s35}
\Sigma_{\mu\nu\Qv}(E_n) =  \sum_{\xi, \alpha \qv} \mathcal{D}_{\alpha \qv}^{{\xi \mu}} \mathcal{D}_{\alpha, -\qv}^{{\nu \xi}} \left( \frac{ \averageo{\hat{n}_{\alpha \qv}}}{iE_n - \Ex_{\xi\Qv-\qv}+ \hbar \Omega_{\alpha \qv}} + \frac{1+\averageo{\hat{n}_{\alpha \qv}}}{iE_n - \Ex_{\xi\Qv-\qv} - \hbar \Omega_{\alpha \qv}} \right),
\end{equation}
which, by analogy with electron–phonon theory, we refer to as the Fan–Migdal self-energy \cite{Giustino2017}. In contrast to the electron–phonon case, which describes the coupling of fermions (electrons) to bosons (phonons), in our microscopic model both interacting particles are bosons: excitons, treated in the low-density limit as bosons, and phonons. Equation \eqref{s34} results from truncating the exponential in Eq. \eqref{s12} at first order. Extending the expansion to higher orders and retaining the terms describing single exciton-phonon interactions (terms of the Fan-Migdal type) leads to the following Dyson equation:
\begin{equation}
\label{s36}
g_{\mu\Qv,\nu\Qv}(E_n) = \delta_{\mu \nu} g_{\mu\Qv, \mu\Qv}^0 (E_n) + g_{\mu\Qv, \mu\Qv}^0(E_n) \sum_{\lambda} \Sigma_{\mu\lambda\Qv}(E_n) g_{\lambda\Qv, \nu\Qv}(E_n),
\end{equation}
or in a compact form
\begin{equation}
\label{s37}
\sum_{\lambda} \left[ (iE_n - \Ex_{\mu\Qv})\delta_{\mu\lambda} - \Sigma_{\mu\lambda\Qv}(E_n)  \right] g_{\lambda\Qv,\nu\Qv}(E_n) =  \delta_{\mu \nu},
\end{equation}
with $\lambda$ labeling exciton states. Performing the standard analytic continuation from discrete imaginary Matsubara frequencies to continuous real frequencies ($iE_n \rightarrow E + i\eta$) yields the following equation for the retarded Green's function $\Gr{\mu\Qv}{\nu\Qv} (E)$:
\begin{equation}
\label{s38}
\sum_{\lambda} \left[ (E - \Ex_{\mu\Qv})\delta_{\mu\lambda} - \Sigma_{\mu\lambda\Qv}(E + i\eta) \right] \Gr{\lambda\Qv}{\nu\Qv} (E) = \delta_{\mu \nu},
\end{equation}
where $\eta$ is a positive infinitesimal number. This expression corresponds directly to Eq. (7) in the main text. To simplify notation, we omit the explicit $+i\eta$ dependence in the Fan–Migdal self-energy and write  $\Sigma_{\mu\lambda\Qv}(E + i\eta) \equiv \Sigma_{\mu\lambda\Qv}(E)$, with the understanding that an additional $+i\eta$ is included in the denominators of Eq. \eqref{s35}.

Lastly, by defining the tensor $M_{\mu\lambda\Qv}(E) =  (E - \Ex_{\mu\Qv})\delta_{\mu\lambda} - \Sigma_{\mu\lambda\Qv}(E)$, Eq. \eqref{s38} can be rewritten as the following product of two matrices
\begin{equation}
\label{s39}
\sum_{\lambda} M_{\mu\lambda\Qv}(E) \Gr{\lambda\Qv}{\nu\Qv} (E) = \delta_{\mu \nu}.
\end{equation}
Expressed in this form, it is clear that the poles of $\Gr{\lambda\Qv}{\nu\Qv}$ corresponding to the polaron energies $\Ep_\Qv$ are given by the roots of the secular equation:
\begin{equation}
\label{s40}
\det(M_{\mu\lambda\Qv}) = \det\left[ (\Ep_\Qv - \Ex_{\mu\Qv})\delta_{\mu\lambda} - \Sigma_{\mu\lambda\Qv}(\Ep_\Qv) \right] = 0.
\end{equation}
This equation is equivalent to Eq. (2) presented in the main text.

\section*{Supplementary Note 2. Numerical calculation details}\label{note2}

To compute the complex polaron energy, $\Ep_\Qv$, the retarded Green's function, $\Gr{\mu\Qv}{\nu\Qv}$, and the corresponding spectral functions, $A_\Qv$, presented in Figs. 2–5 of the main text, we numerically solved Eqs. \eqref{s38}-\eqref{s40}. In this Supplementary Note, we provide details of these numerical simulations.

\subsection*{S2.1 Solution of Dyson (secular) equation for 1\textit{s} exciton polarons}\label{sec2.1}

To obtain the results presented in Figs. 2 and 3 of the main text, we numerically solved Eqs. \eqref{s38} and \eqref{s40}. The system was modeled for $1s\text{ KK}$ excitons coupled to both transverse (TA) and longitudinal (LA) acoustic phonons in monolayer MoSe$_2$. Under these assumptions, Eqs. \eqref{s38} and \eqref{s40} reduce to 
\begin{equation}
\label{s41}
\Gr{1s\Qv}{1s\Qv} (E) = \frac{1}{E - \Ex_{1s,\Qv}- \Sigma_{1s,1s,\Qv}(E)},
\end{equation}
and
\begin{equation}
\label{s42}
\Ep_\Qv - \Ex_{1s,\Qv} - \Sigma_{1s,1s,\Qv}(\Ep_\Qv) = 0.
\end{equation}
Here $\Sigma_{1s,1s,\Qv}$ is the Fan-Migdal self-energy for $1s$ excitons determined by Eq. \eqref{s35}, i.e,
\begin{align}
\label{s43}
\Sigma_{1s,1s,\Qv}(\Ep_\Qv) &= \sum_{\alpha=\text{TA,LA},\qv}\, |\mathcal{D}_{\alpha \qv}^{{1s,1s}}|^2 \left( \frac{ \averageo{\hat{n}_{\alpha \qv}}}{\Ep_\Qv - \Ex_{1s,\Qv-\qv}+ \hbar \Omega_{\alpha \qv}+i\eta} + \frac{1+\averageo{\hat{n}_{\alpha \qv}}}{\Ep_\Qv - \Ex_{1s,\Qv-\qv} - \hbar \Omega_{\alpha \qv}+i\eta} \right).
\end{align}
where the intravalley acoustic exciton-phonon matrix elements $\mathcal{D}_{\alpha \mathbf{q}}^{\mu\nu}$ are computed as
\begin{equation}
\label{s44}
\mathcal{D}^{\mu\nu}_{\alpha\qv} = D^c_{\alpha\qv} \mathcal{F}^{\mu\nu}\left(\frac{m_h}{m_e+m_h}\qv \right) - D_{\alpha \qv}^v \mathcal{F}^{\mu\nu}\left(-\frac{m_e}{m_e+m_h} \qv\right),
\end{equation}
where $D_{\alpha \mathbf{q}}^{c}$ and $D_{\alpha \mathbf{q}}^{v}$ represent the electron- and hole-phonon matrix elements for the TMD monolayer, respectively. These matrix elements are obtained from density functional theory (DFT) calculations as \cite{Kim2014} :
\begin{equation}
\label{s45}
D_{\alpha \qv}^{c(v)} = \sqrt{\frac{\hbar}{2 \rho \mathcal{A} \Omega_{\alpha \qv}}} \tilde{D}_{\alpha \qv}^{c(v)},
\end{equation}
where $\rho = m_\text{uc}/\mathcal{A}$ is the surface mass density of the TMD monolayer, with $m_\text{uc} = m_\text{m} + 2m_\text{c}$ the mass and $\mathcal{A} = \sqrt{3}\,a_0^2/2$ the area of the unit cell. The quantities $m_\text{c}$ and $m_\text{m}$ denote the masses of the chalcogen and transition-metal atoms, respectively, and $a_0$ is the lattice constant of the TMD. The deformation potentials are determined as $\tilde{D}_{\alpha \qv}^{c(v)}= \tilde{D}_{\alpha}^{c(v)} \qv$, and the energy dispersion relation of the acoustic phonons is $\hbar\Omega_{\alpha\qv}= \hbar v_{\alpha} \qv$ with $v_\alpha$ being the velocity of the $\alpha$ acoustic mode. In Eq. \eqref{s44}, the corresponding exciton form factors are defined as $\mathcal{F}^{\mu\nu}(\mathbf{q}) = \sum_{\mathbf{k}} \Psi_{\mu \kv}^{*} \Psi_{\nu \kv+\qv}$ where $\Psi_{\mu\kv}$ denotes the wavefunctions of the exciton obtained by solving the Wannier equation (see the Methods section in the main text). Equations \eqref{s41}-\eqref{s45} also apply to the interaction of $1s$ excitons with optical phonons at the $\Gamma$ point. In this case, we treat the optical phonons in the Einstein approximation, $\hbar\Omega_{\alpha\qv} = \hbar\Omega_\alpha$, with $\alpha$ denoting the three relevant optical modes in the monolayer, i.e., transverse optical (TO), longitudinal optical (LO) and A1 phonon modes \cite{Kim2014}. Following Ref. \cite{Brem2020}, we model the intravalley optical exciton–phonon matrix elements with Eq. \eqref{s44}, reversing the sign of the hole–phonon coupling as $-D_{\alpha\qv}^v \mapsto D_{\alpha\qv}^v$. Numerically solving Eq. \eqref{s42} under these approximations, yields a negligible redshift and renormalized properties of the optical exciton polaron relative to the bare $1s$ exciton. This follows from the reversed sign of the hole–phonon coupling, which causes the electron and hole contributions in Eq. \eqref{s44} to partially cancel. Applying the present microscopic model to describe the optical exciton polaron properties with $\Gamma$-point phonons thus requires a detailed determination of the relative sign of $D_{\alpha\qv}^c$ and $D_{\alpha\qv}^v$, for example by density functional perturbation theory (DFPT) calculations \cite{Baroni2001}. Such calculations are beyond the scope of this work.

We note that solving Eq. \eqref{s42} for the complex polaron energy $\Ep_\Qv$ is equivalent to solving the following system of equations
\begin{subequations}
\label{s46}
\begin{align}
\label{s46a}
\text{Re}(\Ep_\Qv) - \Ex_{1s,\Qv} - \text{Re}[\Sigma_{1s,1s,\Qv}(\Ep_\Qv)] &= 0, \\
\label{s46b}
\text{Im}(\Ep_\Qv) - \text{Im}[\Sigma_{1s,1s,\Qv}(\Ep_\Qv)] &= 0.
\end{align}
\end{subequations}
For momenta $Q > Q_c$, Eq. \eqref{s46b} is no longer satisfied. We thus compute the polaron energy shown in Fig. 2 of the main text by first solving
\begin{equation}
\label{s47}
\text{Re}(\Ep_\Qv) - \Ex_{1s,\Qv} - \text{Re}[\Sigma_{1s,1s,\Qv}(\text{Re}(\Ep_\Qv))] = 0,
\end{equation}
and subsequently setting 
\begin{equation}
\label{s48}
\text{Im}(\Ep_\Qv) = \text{Im}[\Sigma_{1s,1s,\Qv}(\text{Re}(\Ep_\Qv))].
\end{equation}
In contrast, the results shown in Fig. 4 of the main text are derived from the full solution of Eqs. \eqref{s46a} and \eqref{s46b}.

\begin{figure}[t!]
\centering
\includegraphics[width=1.0\linewidth]{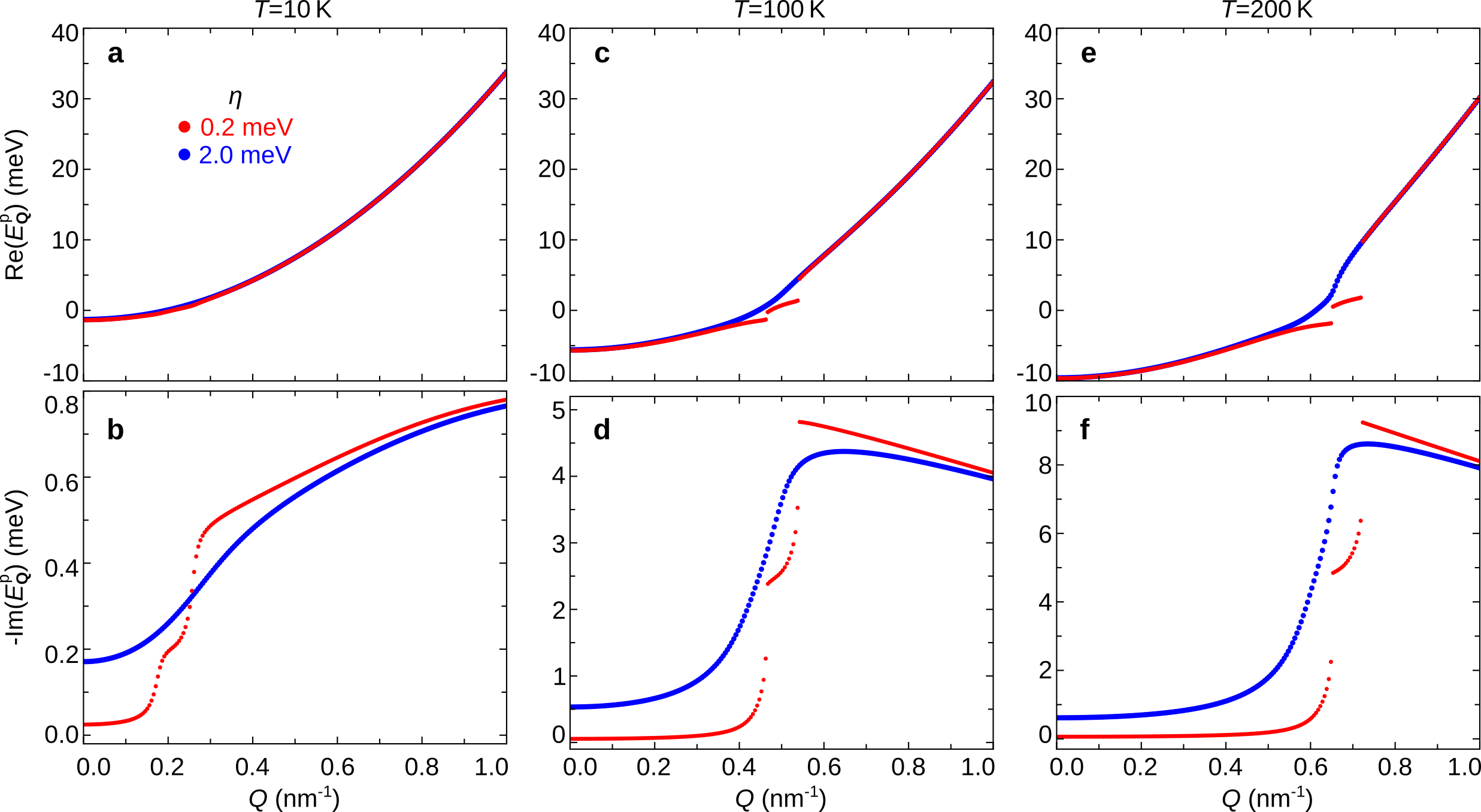}
\caption{\textbf{Complex polaron energies for varying $\boldsymbol{\eta}$.} \textbf{a} Real and \textbf{b} negative imaginary components of the acoustic exciton polaron energy, $\Ep_\Qv$, in a MoSe$_2$ monolayer versus center-of-mass momentum $Q$ at $T = 10\,\text{K}$. Red and blue dots correspond to the results obtained with $\eta=0.2\,\text{meV}$ and $\eta=2\,\text{meV}$, respectively. \textbf{c} -\textbf{d} and \textbf{e}-\textbf{f} are the same as \textbf{a}-\textbf{b} but calculated at $T=100\,\text{K}$ and $200\,\text{K}$. These results are obtained using Eq. \eqref{s47}, and the blue data points in panels c-f are the same dataset presented in Fig. 2 of the main text.}
\label{figS1}
\end{figure}

We assume the thermodynamic limit, and the discrete sums over momentum are converted into continuous two-dimensional integrals as $\sum_\qv \mapsto \frac{1}{(2\pi)^2} \int_0^{\infty} q\, \text{d}q \int_0^{2\pi} \text{d}\theta$. Equations \eqref{s46} and \eqref{s47} are solved via a Newton-Raphson method, and the multidimensional integrals in Eq. \eqref{s43} are evaluated using adaptive cubatures implemented in Julia programming language to ensure numerical convergence. All polaron energies reported in the main text are calculated using $\eta = 2\,\text{meV}$. Including this parameter in the numerical simulations results in an artificial constant offset in the imaginary part of the polaron energy, $\text{Im}(\Ep_\Qv) = -\Gp_\Qv$. This offset is clearly visible in Supplementary Fig. \ref{figS1}, where we show $\Ep_\Qv$ at temperatures of $T = 10\,\text{K}$ (panels a and b), $100\,\text{K}$ (panels c and d), and $200\,\text{K}$ (panels e and f) for the parameters $\eta = 0.2\,\text{meV}$ (red) and $2\,\text{meV}$ (blue). Increasing $\eta$ shifts $\Gp_\Qv$, near $Q \approx 0$, from zero to a non-zero positive value for the three temperatures. Aside from this artificial offset, the polaron dispersion remains identical (compare blue and red dots near $Q \approx 0$ in panels a, c, and e), with a larger $\eta$ serving predominantly to smooth the dispersion curve. Consequently, all renormalized polaron properties near $Q = 0$ derived from the dispersion relations remain independent of the choice of $\eta$.

\subsection*{S2.2 Calculation of the spectral functions}\label{sec2.2}

To compute the spectral function shown in Fig. 3c of the main text, we use the retarded Green's function $\Gr{1s,\Qv}{1s,\Qv}$ defined in Eq. \eqref{s41} and include an additional broadening $\eta$ in its denominator:
\begin{align}
\label{s49}
A_\Qv(E) &= -\frac{1}{\pi} \text{Im}[\Gr{1s\Qv}{1s\Qv}(E+i\eta)] =  -\frac{1}{\pi}  \text{Im} \left[\frac{1}{E - \Ex_{1s,\Qv}- \Sigma_{1s,1s,\Qv}(E) + i \eta}\right] \\
\nonumber
&=\frac{1}{\pi} \frac{\eta - \text{Im}[\Sigma_{1s,1s,\Qv}(E)]}{(E -\Ex_{1s,\Qv} - \text{Re}[\Sigma_{1s,1s,\Qv}(E)])^2+(\eta - \text{Im}[\Sigma_{1s,1s,\Qv}(E)])^2},
\end{align}
This expression has the mathematical form of a Lorentzian function whose full-width at half-maximum (FWHM) can be determined as
\begin{equation}
\label{s50}
\text{FWHM} =  2\, \{\eta - \text{Im}[\Sigma_{1s,1s,\Qv}(E)]\}.
\end{equation}
Using Eq. \eqref{s48}, the full-width at half-maximum (FWHM) evaluated at the polaron energy, $\text{Re}(E) = \text{Re}(\Ep_\Qv)$, takes the form:
\begin{equation}
\label{s51}
\text{FWHM} = 2 (\eta + \Gp_\Qv),
\end{equation}
where $\Gp_\Qv = -\text{Im}[\Sigma_{1s,1s,\Qv}(\text{Re}(\Ep_\Qv))]$ represents the inverse of the intrinsic polaron lifetime. Equation \eqref{s51} establishes the relationship between the FWHM of the spectral function and the polaron lifetime.

To include the intervalley degrees of freedom, and calculate the spectral function shown in Fig. 5 of the main text, we generalize the retarded Green's function from a scalar (Eq. \eqref{s41}) to the following tensor (Eq. \eqref{s39}): 
\begin{equation}
\label{s52}
\Gr{\mu\Qv}{\nu\Qv}(E+i\eta) = [(E +i \eta - \Ex_{\mu\Qv})\delta_{\mu\lambda} - \Sigma_{\mu\lambda\Qv}(E)]^{-1},
\end{equation}
where $\mu$ and $\nu$ run over excitonic states, $\Qv$ is the total center-of-mass momentum and the self-energy $\Sigma_{\mu\lambda\Qv}$ is determined by Eq. \eqref{s35}. For the case of $1s\text{ KK}$ and $1s\text{ KK}'$ excitonic states, the retarded Green's tensor is explicitly
\begin{equation}
\label{s53}
\Gr{\mu\Qv}{\nu\Qv}(E+i\eta) = 
\left[\begin{array}{cc}
E +i\eta - \Ex_{\text{1sKK},\Qv} - \Sigma_{\text{1sKK},\text{1sKK},\Qv}(E) & -\Sigma_{\text{1sKK},\text{1sKK}',\Qv}(E) \\
-\Sigma_{\text{1sKK}',\text{1sKK},\Qv}(E) & E+ i\eta - \Ex_{\text{1sKK}',\Qv} - \Sigma_{\text{1sKK}',\text{1sKK}',\Qv}(E)
\end{array}\right]^{-1}.
\end{equation}
Extending the definition of Eq. \eqref{s44}, the intervalley exciton-phonon matrix elements $\mathcal{D}_{j \qv\Qv}^{\mu\nu}$ entering the self-energies $\Sigma_{\mu\nu\Qv}$ (Eq. \eqref{s35}) are generalized to the expression presented in Ref. \cite{Meneghini2025}:
\begin{equation}
\label{s54}
\mathcal{D}^{\mu\nu}_{j \qv\Qv} = D^{\xi^{\mu}_e, \xi^{\nu}_e, c}_{j\qv} \delta_{\xi^{\mu}_h, \xi^{\nu}_h} \delta_{\xi^{\nu}_e-\xi^{\mu}_e,\xi^{j}_{\text{ph}}} \mathcal{F}^{\mu\nu}\left(\frac{m^h_\nu}{\mx_\nu} \left[\qv + s_{\mu\nu} \Qv \right]\right) -  D^{\xi^{\mu}_h, \xi^{\nu}_h, v}_{j\qv} \delta_{\xi^{\mu}_e, \xi^{\nu}_e} \delta_{\xi^{\nu}_h-\xi^{\mu}_h,\xi^{j}_{\text{ph}}} \mathcal{F}^{\mu\nu}\left(-\frac{m^e_\nu}{\mx_\nu} \left[\qv + s_{\mu\nu} \Qv \right]\right),
\end{equation}
where $j=(\kappa_j,\xi^{\text{ph}}_j)$ is a compound mode index labeling the phonon mode $\kappa_j$ and phonon valley $\xi^{\text{ph}}_j$. Furthermore, $\xi^{\mu}_e$ ($\xi^{\mu}_h$) specifies the high-symmetry valley for the electron (hole) with mass $m^e_\mu$ ($m^h_\mu$); $\mx_\mu = m^e_\mu + m^h_\mu$ is the mass of the exciton, and $s_{\mu\nu}= 1- \mx_\nu/\mx_\mu$ is the mass imbalance. For the range of center-of-mass momenta considered in this work and the exciton masses of both MoSe$_2$ and WSe$_2$, $s_{\mu\nu}\Qv$ is typically a small vector leading to small corrections. Consequently, to reduce the computational time in the calculation of the intervalley exciton-phonon matrix elements (Eq. \eqref{s54}), we neglect this term and approximate $\mathcal{D}^{\mu\nu}_{j \qv\Qv} \approx \mathcal{D}^{\mu\nu}_{j \qv,\Qv=0}\equiv \mathcal{D}^{\mu\nu}_{j \qv} $. The generalized electron- and hole-phonon matrix elements $D^{\xi^{\mu}_e, \xi^{\nu}_e, c}_{j\qv}$ and $D^{\xi^{\mu}_h, \xi^{\nu}_h, v}_{j\qv}$, respectively, are obtained from DFT calculations as \cite{Kim2014} : 
\begin{equation}
\label{s55}
D^{\xi^{\mu}_\lambda, \xi^{\nu}_\lambda, \lambda}_{j\qv} = \sqrt{\frac{\hbar}{2 \rho \mathcal{A} \Omega_{j \qv}}} \tilde{D}_{j\qv}^{\xi^{\mu}_\lambda, \xi^{\nu}_\lambda, \lambda},
\end{equation}
with
\begin{equation}
\label{s56}
\tilde{D}_{j\qv}^{\xi^{\mu}_\lambda, \xi^{\nu}_\lambda, \lambda} = \left\{
\begin{array}{l}
\tilde{D}_{\xi}^{\lambda} \, \qv \hspace{3mm} \text{if} \hspace{2mm}  \xi^{\mu}_\lambda = \xi^{\nu}_\lambda = \xi \hspace{2mm}  \text{and} \hspace{2mm}  \kappa_j = \text{TA, LA} \\
\tilde{D}_{\xi^{\mu}_\lambda \xi^{\nu}_\lambda}^{\lambda} \hspace{3mm} \text{else}
\end{array}\right.,
\end{equation}
and
\begin{equation}
\label{s57}
\Omega_{j\qv} = \left\{
\begin{array}{l}
v_j \qv \hspace{3mm} \text{if} \hspace{2mm} \kappa_j = \text{TA, LA} \\
\Omega_j \hspace{3mm} \text{else}
\end{array}\right..
\end{equation}
Here $\lambda=c,v$ labels the electronic band index, and ``else'' denotes optical modes and intervalley contributions. In Eq. \eqref{s57}, to model the phonon energies, we employ a linear dispersion (Debye
approximation) for acoustic modes near the $\Gamma$ point, whereas a constant energy (Einstein approximation) is used for short-wavelength acoustic phonons ($\xi \neq \Gamma$). Once the intervalley exciton-phonon matrix elements and the self-energies are computed, we insert them into the retarded Green's tensor (Eq. \eqref{s53}) and calculate the spectral function shown in Fig. 5 of the main text, following Ref. \cite{Huang2023}:
\begin{equation}
\label{s58}
A_{\Qv}(E) = -\frac{1}{\pi} \text{Im}\{\text{Tr}[\Gr{\mu\Qv}{\nu\Qv}(E+i\eta)]\}. 
\end{equation}
Within the calculation of $A_{\Qv}(E)$, we set $\eta= 2\,\text{meV}$.

\begin{figure}[t!]
\centering
\includegraphics[width=1.0\linewidth]{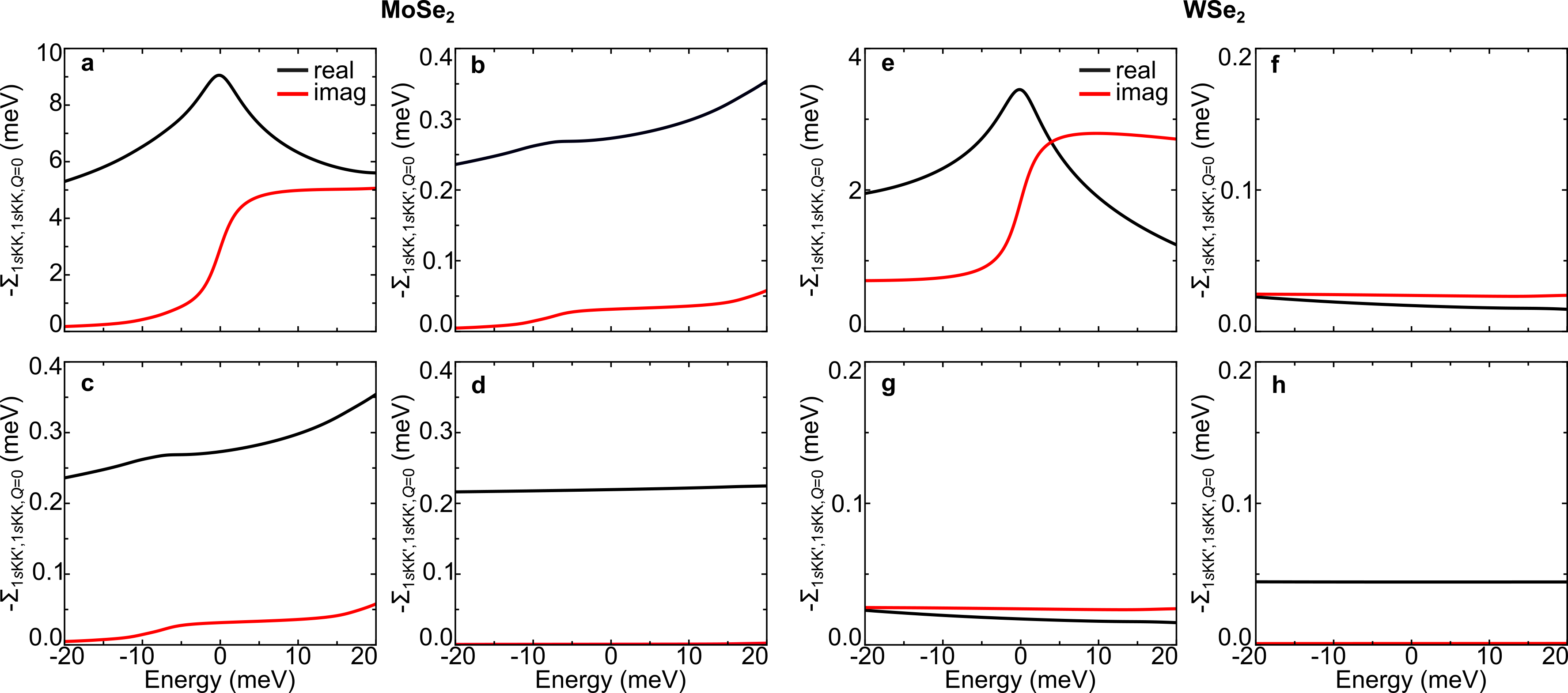}
\caption{\textbf{Intervalley self-energies $\boldsymbol{\Sigma_{\mu\nu,\Qv}}$ in MoSe$_2$ and WSe$_2$.} \textbf{a}-\textbf{d} Real (black) and imaginary (red) components of the self-energy matrix elements $\Sigma_{\mu\nu,\Qv}$ with $\mu,\nu \in \{1s\text{KK}, 1s\text{KK}'\}$ for MoSe$_2$, evaluated at
$Q = 0$ and $T = 100\,\text{K}$. Panels \textbf{a} and \textbf{d} show the diagonal and panels
\textbf{b} and \textbf{c} the off-diagonal elements. \textbf{e}-\textbf{h} Same as
\textbf{a}-\textbf{d}, but for WSe$_2$. The horizontal energy axes are defined relative to $\Ex_{1s\text{KK},\Qv=0}$ of the respective material.}
\label{figS2}
\end{figure}

Supplementary Fig. \ref{figS2} shows the real (black lines) and imaginary (red lines) components of the intervalley self-energies entering Eq.~\eqref{s53}, evaluated at $Q = 0$ and $T = 100\,\text{K}$. These self-energies are used to compute the spectral functions in Figs. 5b and 5f of the main text according to Eq. \eqref{s58}. For MoSe$_2$ (panels a-d), the off-diagonal elements (panels b and c) are dominated by their real parts and are not negligible. When the matrix is inverted to obtain the spectral function, they enter through the product $\Sigma_{1s\text{KK},1s\text{KK}',0}\Sigma_{1s\text{KK}',1s\text{KK},0}$, which couples the KK and KK' states and produces a sizable shift of the polaron energy. In contrast, for WSe$_2$ (panels e-h), the real and imaginary components of the off-diagonal elements (panels f and g) are comparable and much smaller than the diagonal components (panels e and h), so the coupling mainly broadens the resonance rather than shifting it.

\subsection*{S2.3 Group velocity, effective mass and lifetime of exciton polarons}\label{sec2.3}

To evaluate the renormalized properties of the exciton polarons shown in the main text, we use the dispersion relation of the polaron obtained from Eqs. \eqref{s46a}-\eqref{s46b}  and calculate the polaron group velocity via the first numerical derivative of the dispersion relation, $v_{\mathbf{Q}}^{\text{p}} = \nabla_\Qv \text{Re}(\Ep_\Qv)/\hbar$, while the effective mass is computed from the second derivative as $\mpol = \hbar^2 / \nabla^2_\Qv \text{Re}(\Ep_\Qv)$. To extract the polaron lifetime from the spectral function, we first obtain the full width at half maximum of the spectra and compute $\Gp_\Qv = \text{FWHM}/2 - \eta$, where $\eta = 2\,\text{meV}$ is the broadening parameter introduced in the calculation. The lifetime is then given by $\tau_\Qv = \hbar/(2\Gp_\Qv)$.

\clearpage

\subsection*{S2.4 Material parameters used in the numerical simulations}\label{sec2.4}

Supplementary Tables \ref{table1}-\ref{table2} show all material parameters used in the numerical simulations.

\begin{table}[h!]
\centering
\begin{tabular}{ccc}
\hline\hline
Parameter & MoSe$_2$ & WSe$_2$  \\
\hline
\small{$a_0$ (nm)}          & \small{0.335}      & \small{0.334}  \\
\small{$d_0$ (nm)}          & \small{0.645}      & \small{0.652} \\
\small{$m_\text{c}$ (u)}    & \small{95.0}       & \small{183.0} \\
\small{$m_\text{m}$ (u)}    & \small{79.0}       & \small{79.0} \\
$\varepsilon_{\parallel}$   & \small{16.5}       & \small{15.1} \\
$\varepsilon_{\bot}$        & \small{7.4}        & \small{7.5} \\
$\varepsilon_{\text{hBN}}$  & \small{4.5}        & \small{4.5} \\
\small{$m^e_{\text{K}}$ ($m_0$)}         & \small{0.50}        & \small{0.29} \\
\small{$m^h_{\text{K}}$ ($m_0$)}         & \small{0.60}        & \small{0.36} \\
\small{$m^e_{\text{K}'}$ ($m_0$)}        & \small{0.58}        & \small{0.40} \\
\small{$m^h_{\text{K}'}$ ($m_0$)}        & \small{0.70}        & \small{0.54} \\
\small{$E_{\text{bg}}$ (eV)}             & \small{1.5604}      & \small{1.6910} \\
\small{$\Delta^e_{\text{KK}'}$ (meV)}    & \small{22}          & \small{-37} \\
\hline
\end{tabular}
\caption{\textbf{Parameters of the TMD mononlayer encapsulated in hBN.} Dielectric parameters are taken from Ref. \cite{Laturia2018}, and the atomic masses of the transition metal and chalcogen atoms from Ref. \cite{NIST}. Effective masses of electrons and holes near the high-symmetry points K and K$'$ are given in units of the free electron mass $m_0$. The energy gap between the K valley in the valence band and the K$'$ valley in the conduction band is obtained as $E^{\text{g}}_{\text{KK}'} = E_{\text{bg}} + \Delta^{e}_{\text{KK}'}$, where $E_{\text{bg}}$ is the direct bandgap energy and $\Delta^{e}_{\text{KK}'}$ is the energy difference between the K and K$'$ valleys in the conduction band. The effective masses, $E_{\text{bg}}$, and $\Delta^{e}_{\text{KK}'}$ are taken from Ref. \cite{Kormanyos2015}.}
\label{table1}
\end{table}

\begin{table}[h!]
\centering
\begin{tabular}{ccc}
\hline\hline
Parameter & MoSe$_2$ & WSe$_2$  \\
\hline
\small{$v_\text{TA/LA}$ (nm/fs)}   & \small{$4.1\times10^{-3}$}  & \small{$3.3\times10^{-3}$} \\
%
\small{$\hbar \Omega_{\text{TO}, \Gamma}$ (meV)} & \small{36.1}   & \small{30.5} \\
%
\small{$\hbar \Omega_{\text{LO}, \Gamma}$ (meV)} & \small{36.6}   & \small{30.8} \\
%
\small{$\hbar \Omega_{\text{A}_1, \Gamma}$ (meV)} & \small{30.3}   & \small{30.8} \\
%
\small{$\hbar \Omega_{\text{TA}, \text{K}}$ (meV)} & \small{16.6}   & \small{15.6} \\
%
\small{$\hbar \Omega_{\text{LA}, \text{K}}$ (meV)} & \small{19.9}   & \small{18.0} \\
%
\small{$\tilde{D}^{c}_{\text{ac},\Gamma}$ (eV)} & \small{$3.4/\sqrt{2}$}   & \small{$3.2/\sqrt{2}$}  \\
%
\small{$\tilde{D}^{v}_{\text{ac},\Gamma}$ (eV)} & \small{$2.8/\sqrt{2}$}   & \small{$2.1/\sqrt{2}$}  \\
%
\small{$\tilde{D}^{c}_{\text{op},\Gamma}$ (eV/nm)} & \small{52.0}   & \small{23.0}  \\
%
\small{$\tilde{D}^{v}_{\text{op},\Gamma}$ (eV/nm)} & \small{49.0}   & \small{31.0}  \\
%
\small{$\tilde{D}^{c}_{\text{ac, K}}$ (eV/nm)} & \small{18.0}   & \small{13.0}  \\
%
\small{$\tilde{D}^{v}_{\text{ac, K}}$ (eV/nm)} & \small{9.0}   & \small{11.0}  \\
\hline
\end{tabular}
\caption{\textbf{Phonon parameters.}    Summary of the parameters used for phonon velocities, phonon energies, and deformation potentials taken from Ref. \cite{Kim2014}.}
\label{table2}
\end{table}


\section*{Supplementary Note 3. Branch points and branch cuts in the retarded Green's function}\label{note3}

\begin{figure}[H]
\centering
\includegraphics[width=0.82\linewidth]{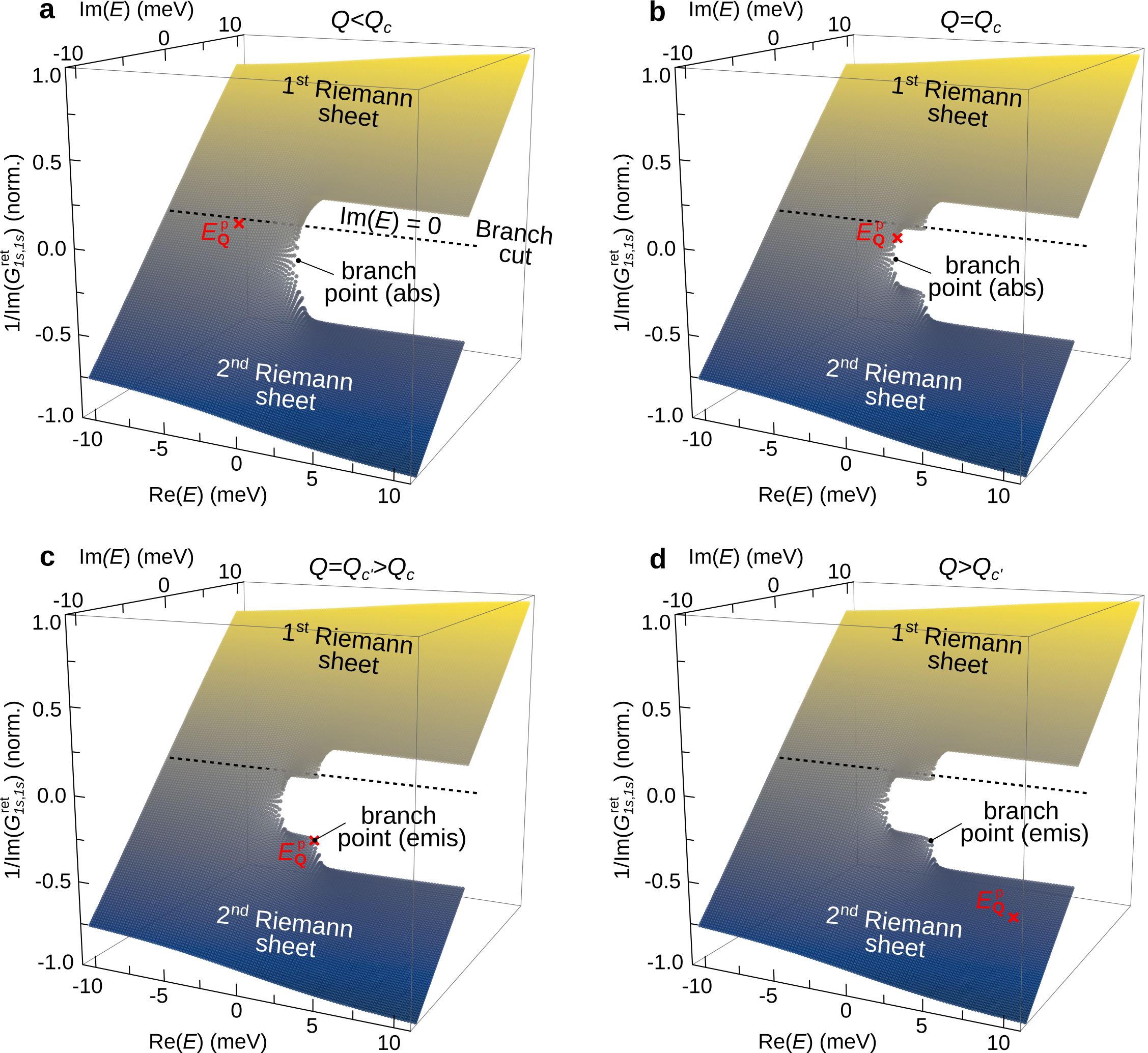}
\caption{\textbf{Branch points and branch cuts}. \textbf{a}-\textbf{d} Normalized 3D plots of the inverse imaginary part of the retarded Green's function, $1/\text{Im}[\Gr{1s\Qv}{1s\Qv}(E)]$, as a function of complex energy $E$ at $T=100\,\text{K}$ for MoSe$_2$. The contour plots are calculated at $Q=0.1\,\text{nm}^{-1}$ (panel a), $Q=Q_c=0.407\,\text{nm}^{-1}$ (panel b), $Q = Q_{c'}=0.48\,\text{nm}^{-1}$ (panel c) and $Q=0.6\,\text{nm}^{-1}$ (panel d). The blue-to-yellow color gradient on the surface indicates the varying $z$-axis heights of the data points. The black dashed lines marks the condition $\text{Im}(E)=0$, while the red crosses indicate the polaron energy $\Ep_\Qv$, i.e., the solution of Eqs. \eqref{s47}-\eqref{s48} at the corresponding values of $Q$. Labels ``abs'' and ``emis'' mark the branch points in $1/\text{Im}[\Gr{1s\Qv}{1s\Qv}]$. The asymmetry of $1/\text{Im}[\Gr{1s\Qv}{1s\Qv}]$ along the dashed line is a consequence of the finite broadening parameter $\eta=2\,\text{meV}$. The energy axes in all panels are defined relative to $\Ex_{1s,Q=0}$ and $1/\text{Im}[\Gr{1s\Qv}{1s\Qv}]$ is normalized to its maximum value.}
\label{figS3}
\end{figure}

In this Supplementary Note, we show that the kinks in the polaron dispersion emerge where the polaron energy $\Ep_\Qv$ reaches a branch point of the retarded Green's function (Eq. \eqref{s41}), i.e., the onset of the continuum of exciton-phonon scattering. To this end, we focus on the representative temperature $T = 100\,\text{K}$ and show in Fig. \ref{figS3} three-dimensional surface plots of the inverse imaginary part of the retarded Green's function, $1/\text{Im}[\Gr{1s\Qv}{1s\Qv}(E)]$. These plots allow us to follow the trajectory of the polaron energy $\Ep_\Qv$ (red crosses) in the complex energy plane as a function of the center-of-mass momentum $Q$.

For $Q < Q_c$ (panel a), $\Ep_\Qv$ lies on the first (physical) Riemann sheet and corresponds to a bound polaron state with a purely real energy. The slight displacement of the red cross below the real axis (black dashed line) is due to the finite broadening $\eta = 2\,\text{meV}$ introduced in the calculation, which shifts the pole slightly off the real axis. As $Q$ approaches $Q_c$ (panel b), $\Ep_\Qv$ moves toward the acoustic-phonon absorption branch point (labeled ``abs'') of $\Gr{1s\Qv}{1s\Qv}$, which defines the energy threshold for acoustic-phonon absorption. For $Q > Q_c$ (panel c), the pole crosses the associated branch cut onto the second Riemann sheet and acquires a negative imaginary component beyond the offset caused by $\eta$, moving downward from the real axis. Upon further increasing $Q$ (panel d), the pole approaches a second branch point of the retarded Green's function, associated with acoustic-phonon emission (labeled ``emis''). At a second critical momentum $Q = Q_{c'}$, $\Ep_\Qv$ aligns with this emission threshold. For $Q > Q_{c'}$, the emission channel opens as well, and the negative imaginary part of the pole increases further, which physically manifests as a finite, and shorter, quasiparticle lifetime.

\section*{Supplementary Note 4. Spectral function of intervalley exciton polarons in MoSe$_2$ and WSe$_2$}\label{note4}

\begin{figure}[H]
\centering
\includegraphics[width=0.82\linewidth]{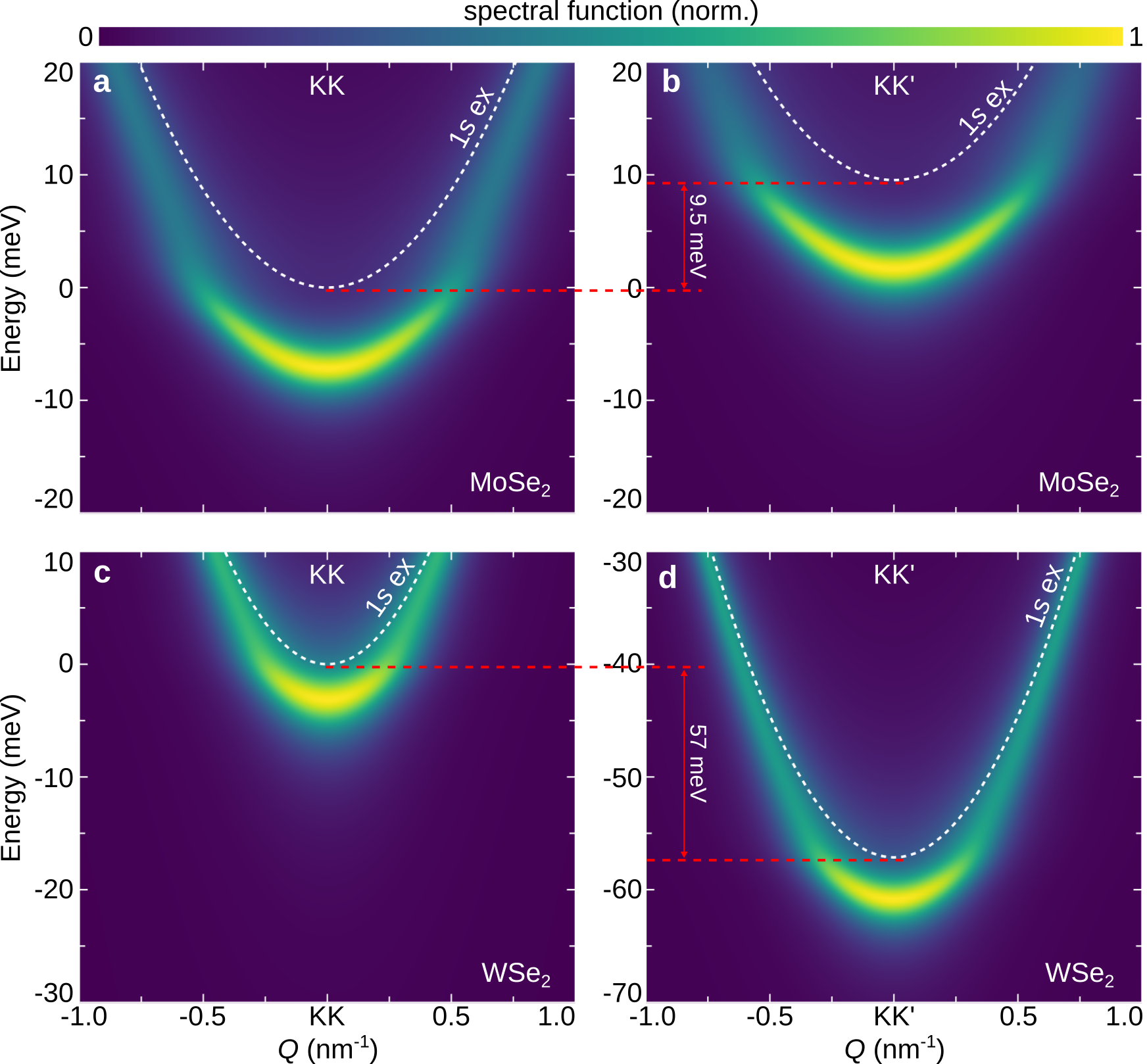}
\caption{\textbf{Spectral function of KK and KK' exciton polarons in MoSe$_2$ and WSe$_2$}. \textbf{a} Calculated spectral function $A_\Qv(E)$ as a function of momentum $Q$ and energy $E$ for the $1s \text{ KK}$ acoustic exciton polaron in a MoSe$_2$ monolayer, incorporating coupling to the $1s \text{ KK}'$ intervalley exciton via scattering with acoustic K-point phonons. \textbf{b} Same as \textbf{a}, but for the energies and momenta in the vicinity of the $1s \text{ KK}'$ exciton. White dashed curves indicate the dispersion relations of the $1s \text{ KK}$ (panel a) and the $1s \text{ KK}'$ (panel b) excitons in MoSe$_2$. Red horizontal lines and a double-headed arrow indicate the 9.6\,meV energy separation between the $1s\text{ KK}$ and $1s\text{ KK}'$ exciton states at $Q=0$. \textbf{c}-\textbf{d} Same as \textbf{a}-\textbf{b} but for a WSe$_2$ monolayer. Energy axes are defined relative to $\Ex_{1s\text{KK}, \Qv=\bf{0}}$ in MoSe$_2$ (panels a and b) and WSe$_2$ (panels c and d). All results are calculated at $T=100\,\text{K}$ and normalized to their maximum value. Panels a and c reproduce the data shown in Figs 5a and 5e of the main text, respectively.}
\label{figS4}
\end{figure}

In Supplementary Figs. \ref{figS4}a and b we show the spectral function of an MoSe$_2$ monolayer at $T = 100\,\text{K}$ as a function of energy and momentum in the vicinity of the $1s\text{ KK}$ (panel a) and $1s\text{ KK}'$ (panel b) excitons. Supplementary Figs. \ref{figS3}c and \ref{figS3}d show the same for a WSe$_2$ monolayer. The white dashed curves show the dispersion of the corresponding $1s$ exciton. In MoSe$_2$, the $1s\text{ KK}'$ exciton lies $9.6\,\text{meV}$ above the $1s\text{ KK}$ exciton. Scattering into the KK' valley therefore requires the absorption of a K-point phonon, which is suppressed by the low thermal occupation of these phonons ($\hbar\Omega_\text{TA, K} = 16.6\,\text{meV}$ and $\hbar\Omega_\text{LA, K} = 19.9\,\text{meV}$). Consequently, intervalley coupling does not broaden the $1s\text{ KK}$ polaron resonance, whose linewidth near $Q =0$ is comparable to that of the $1s\text{ KK}'$ resonance (see panels a and b). In contrast, in WSe$_2$ the $1s\text{ KK}'$ exciton lies $57\,\text{meV}$ below the $1s\text{ KK}$ exciton, allowing $1s\text{ KK}$ excitons to scatter into $1s\text{ KK}'$ states by spontaneously emitting K-point phonons. Since the $1s\text{ KK}$ exciton energy increases with $Q$, the emission scattering channel remains open at all momenta and broadens the $1s\text{ KK}$ polaron resonance across its the entire dispersion (see panels c and d).

\section*{Supplementary Note 5. Energy shift and linewidth of exciton polarons}\label{note5}

In Supplementary Fig. \ref{figS5}, we show the spectral functions of MoSe$_2$ and WSe$_2$ as a function of temperature, as discussed in the main text. Panels c, d, g and h are identical to Figs. 5c, 5d, 5g, and 5h of the main text.

\begin{figure}[H]
\centering
\includegraphics[width=0.9\linewidth]{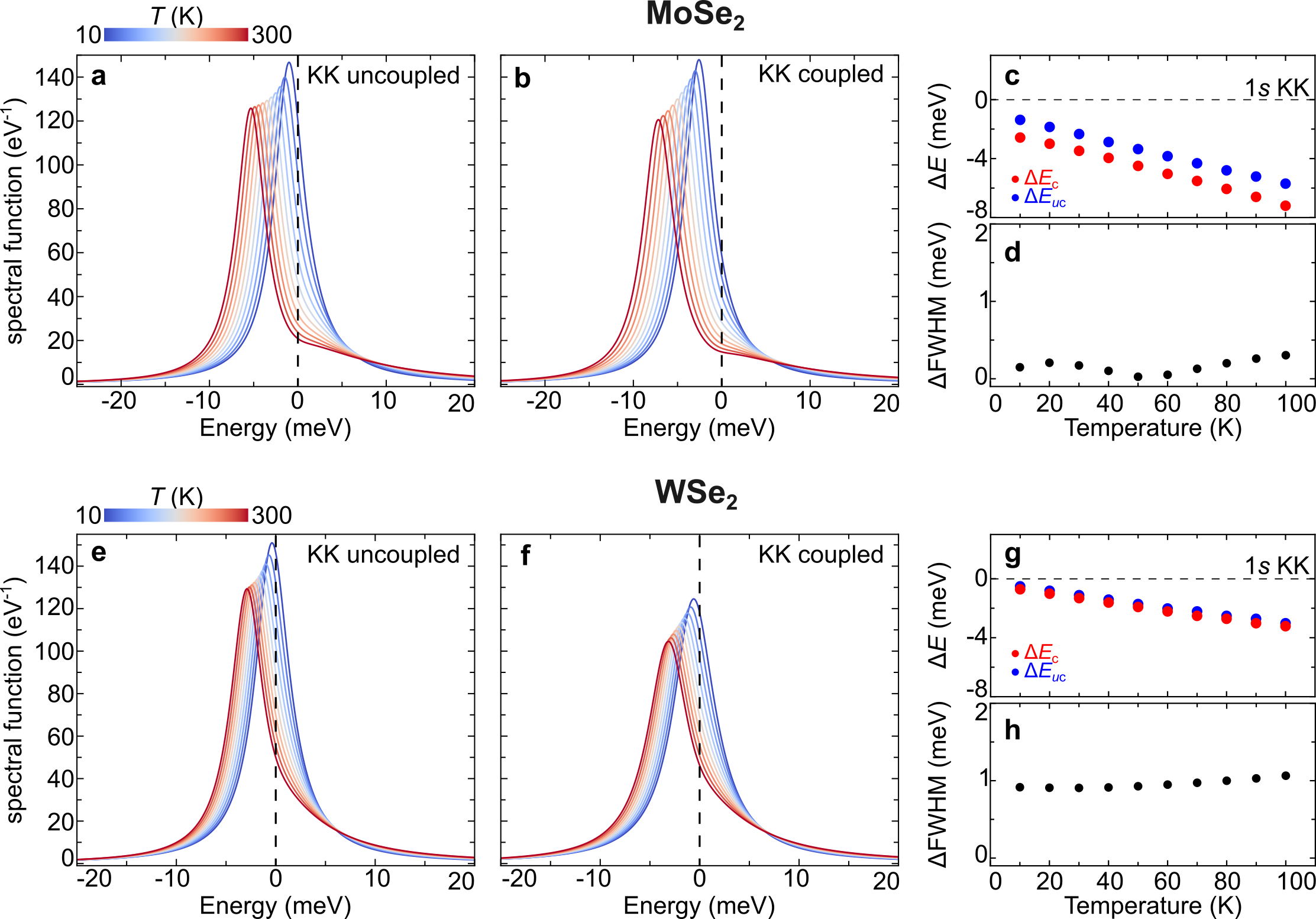}
\caption{\textbf{Temperature-dependent spectral function in MoSe$_2$ and WSe$_2$}. \textbf{a} Calculated spectral function $A_\Qv(E)$ as a function of energy for the $1s\text{ KK}$ acoustic exciton polaron in an MoSe$_2$ monolayer in the absence of coupling to the $1s\text{ KK}'$ intervalley exciton, evaluated at temperatures from $10\text{ K}$ to $100\text{ K}$ in $10\text{ K}$ increments. The spectral functions are calculated at $Q = 0$ and the vertical dashed line marks $\Ex_{1s\text{KK},\Qv=\bf 0}$.
\textbf{b} Same as in \textbf{a}, but incorporating coupling to the $1s\text{ KK}'$ intervalley exciton via acoustic K-point phonon scattering.
\textbf{c} Polaron energy redshift $\Delta E$ as a function of temperature for the coupled (red dots) and uncoupled (blue dots) systems, referenced to $\Ex_{1s\text{KK}, \Qv=\bf 0}$ indicated by the horizontal dashed line. \textbf{d} Difference in spectral linewidth between coupled and uncoupled systems, $\Delta\text{FWHM} = \vert\text{FWHM}_{\text{c}} - \text{FWHM}_{\text{uc}}\vert$, as a function of temperature for the $1s\text{ KK}$ state. \textbf{e}-\textbf{h} Same as \textbf{d}-\textbf{f}, but for a WSe$_2$ monolayer. All spectra are normalized to their peak maximum.}
\label{figS5}
\end{figure}


%